\documentclass[twocolumn,showpacs,preprintnumbers,amsmath,amssymb,pra,superscriptaddress]{revtex4}
\usepackage[latin1]{inputenc}
\usepackage{dcolumn,color}
\usepackage{bm}
\usepackage{graphicx}
\usepackage[english]{babel}

\DeclareMathOperator{\Tr}{Tr}
\DeclareMathOperator{\Rea}{Re}
\DeclareMathOperator{\Ima}{Im}
\def\bbm[#1]{\mbox{\boldmath $#1$}}
\newcommand{\ket}[1]{\displaystyle{|#1\rangle}}
\newcommand{\bra}[1]{\displaystyle{\langle #1|}}
\newcommand{\TE}{\text{TE}}
\newcommand{\TM}{\text{TM}}
\newcommand{\red}[1]{{\color{red}#1}}

\graphicspath{{./Figures/}}

\begin{document}

\title{Three-body radiative heat transfer and Casimir-Lifshitz force\\out of thermal equilibrium for arbitrary bodies}

\author{Riccardo Messina}\email{riccardo.messina@univ-montp2.fr}
\affiliation{Universit\'{e} Montpellier 2, Laboratoire Charles Coulomb UMR 5221 - F-34095, Montpellier, France}\affiliation{CNRS, Laboratoire Charles Coulomb UMR 5221 - F-34095, Montpellier, France}
\author{Mauro Antezza}\email{mauro.antezza@univ-montp2.fr}
\affiliation{Universit\'{e} Montpellier 2, Laboratoire Charles Coulomb UMR 5221 - F-34095, Montpellier, France}\affiliation{CNRS, Laboratoire Charles Coulomb UMR 5221 - F-34095, Montpellier, France}
\affiliation{Institut Universitaire de France - 103, bd Saint-Michel - F-75005 Paris, France}

\date{\today}

\begin{abstract}
We study the Casimir-Lifshitz force and the radiative heat transfer in a system consisting of three bodies held at three independent temperatures and immersed in a thermal environment, the whole system being in a stationary configuration out of thermal equilibrium. The theory we develop is valid for arbitrary bodies, i.e. for any set of temperatures, dielectric and geometrical properties, and describes each body by means of its scattering operators. For the three-body system we provide a closed-form unified expression of the radiative heat transfer and of the Casimir-Lifshitz force (both in and out of thermal equilibrium). This expression is thus first applied to the case of three planar parallel slabs. In this context we discuss the non-additivity of the force at thermal equilibrium, as well as the equilibrium temperature of the intermediate slab as a function of its position between two external slabs having different temperatures. Finally, we consider the force acting on an atom inside a planar cavity. We show that, differently from the equilibrium configuration,  the absence of thermal equilibrium admits one or more positions of minima for the atomic potential. While the corresponding  atomic potential depths are very small for typical ground state atoms, they may become particularly relevant for Rydberg atoms, becoming a promising tool to produce an atomic trap.
\end{abstract}


\maketitle

\section{Introduction}

The quantum and classical fluctuations of the electromagnetic field are at the origin of several physical phenomena such as the existence of a force, even in vacuum (the ground state of electromagnetic field), between any couple of polarizable bodies. This effect, usually known as Casimir-Lifshitz effect between macroscopic bodies and Casimir-Polder force when one or more atoms are involved, was first theoretically predicted in 1948 by Casimir and Polder \cite{CasimirProcKNedAkadWet48,CasimirPhysRev48} for ideal reflecting bodies at $T=0$, and later extended to real bodies at non-zero temperature by Dzyaloshinskii, Lifshitz, and Pitaevskii \cite{DzyaloshinskiiAdvPhys61}.
Recently their predictions have been experimentally verified for several different geometrical configurations \cite{CasimirDalvit}. While these forces have been typically studied at thermal equilibrium, it was shown in 2005 that systems out of thermal equilibrium show indeed new qualitative features, such as a strong tunability of the force and the possibility of switching from an attractive to a repulsive behavior \cite{AntezzaPRL05,AntezzaJPhysA06}. These theoretical predictions were at the origin of the first measurement of the temperature dependence of Casimir force \cite{ObrechtPRL07} by using a BEC of rubidium atoms as a micro-mechanical sensor of force \cite{AntrezzaPRA04}.
These results triggered a new interest in the study of Casimir interactions in several different non-equilibrium scenarios including both body-body \cite{AntezzaPRL06,AntezzaPRA08} and atom-body \cite{BuhmannPRL08,SherkunovPRA09,BehuninPRA10,BehuninJPhysA10,BehuninPRA11} configurations.

The absence of thermal equilibrium is at the origin of a more familiar physical phenomenon also originating from the electromagnetic field fluctuations, namely the radiative heat transfer \cite{VolokitinRevModPhys07}. Remarkably, this effect shares a close theoretical formalism with Casimir-Lifshitz forces, and has also been recently experimentally assessed in different geometries \cite{KittelPRL05,HuApplPhysLett08,NarayanaswamyPRB08,RousseauNaturePhoton09,ShenNanoLetters09,KralikRevSciInstrum11,OttensPRL11,vanZwolPRL12a,vanZwolPRL12b,KralikPRL12}.

In last years, a number of theories have been developed to describe Casimir force and heat transfer out of thermal equilibrium between bodies with arbitrary geometries and dielectric properties \cite{BimontePRA09,MessinaEurophysLett11,MessinaPRA11,KrugerPRL11,KrugerEurophysLett11,KrugerPRB12,RodriguezPRL11,McCauleyPRB12,RodriguezPRB12,RodriguezPRB13}. These theories are based on different approaches (such as scattering matrices, Green's functions and fluctuating surface currents) but all share the use of flucutation-dissipation theorem as the main tool to describe the correlations of the electromagnetic field radiated by each body out of thermal equilibrium. Some of these theories \cite{KrugerPRL11,RodriguezPRB13} have been presented in a general way in order to be able to deal with the general problem of an arbitrary number of bodies. Nevertheless, closed form expressions and numerical applications have been presented only for two interacting bodies.

Non-equilibrium studies dealt also with several applications, as for the case of the heat transfer between two nanogratings \cite{LussangePRB12,GueroutPRB12}. More recently, the absence of equilibrium has been proposed as a tool to manipulate the quantum state of an atom when placed in proximity of a body, realizing a new cooling mechanism and the inversion of the atomic populations \cite{BellomoEurophysLett12,BellomoPRA13} . It has been also suggested to exploit non-equilibrium configurations to produce and protect steady entanglement for two quits \cite{BellomoEurophysLett13,BellomoNewJPhys13}. The heat transfer in configurations involving more than two bodies has been also investigated: heat transfer for three nanoparticles in the dipole approximation \cite{BenAbdallahPRL11}, the time-dependent heating and cooling in a system of an arbitrary number of dipoles in \cite{MessinaPRB13}, the heat transfer in a collection of nanoparticles interacting  with an external laser source \cite{YannopapasPRL13},  and genuine three-body effect proposed to amplify the heat-transfer in a set of three parallel slabs \cite{MessinaPRL12}. Concerning the force, only the three-body force for three atoms at thermal equilibrium has been calculated \cite{PassanteJPhysB98,RizzutoPRL07}.

In this work we develop in detail a general approach to the Casimir force out of thermal equilibrium and heat transfer in a system of three bodies immersed in an environment. The theory we present is valid for arbitrary dielectric and geometrical properties of the body, as well as for arbitrary values of the four temperatures involved (the three temperatures of the bodies and the environmental one). As a side result we obtain the three-body expression for the Casimir-Lifshitz force at thermal equilibrium  for arbitrary bodies. As for the theory for two bodies first presented in \cite{MessinaPRA11}, of which this work is a generalization, in our work we take into account the properties of each body independently by means of its scattering (reflection and transmission) operators. This technique avoids the need to tackle for a given configuration the entire electromagnetic problem, as in any Green-function formulation.

The paper is organized as follows. In Sec. \ref{SecPhysSyst} we present the physical system and the main definitions. In Sec. \ref{SecT} we define the Maxwell stress tensor and the Poynting vector, the main ingredients to calculate the Casimir force and the heat transfer respectively. Section \ref{SecCorr} contains the derivation of the correlation functions of the total field in any region, based on the knowledge of the correlation functions of the source fields as well as on the introduction of the scattering operators. This allows us to give, in Sec. \ref{SecGenFlux}, the flux of Maxwell stress tensor and Poynting vector in a unified formulation. We deal with the case of thermal equilibrium in Sec. \ref{SecThEq}, while Sec. \ref{SecThNeq} contains the main result of our paper, i.e. the expression of the force and the heat transfer for three arbitrary bodies out of thermal equilibrium. In Sec. \ref{Sec3s} this formula is specialized to the case of three parallel slabs. In this part we present as a numerical application a quantitative study of non-additivity of the force acting on one of the external slabs at thermal equilibrium, as well as a study of the equilibrium temperature of the intermediate slab for a given set of the other three temperatures. In Sec. \ref{SecAt} we consider the case of an atom between two parallel slabs, by showing how thermal non-equilibrium allows to design the shape of the force acting on the atom. We finally provide some conclusive remarks in Sec. \ref{SecConcl}.

\section{Physical system and electromagnetic field}\label{SecPhysSyst}

\begin{figure}[htb]
\includegraphics[height=3.3cm]{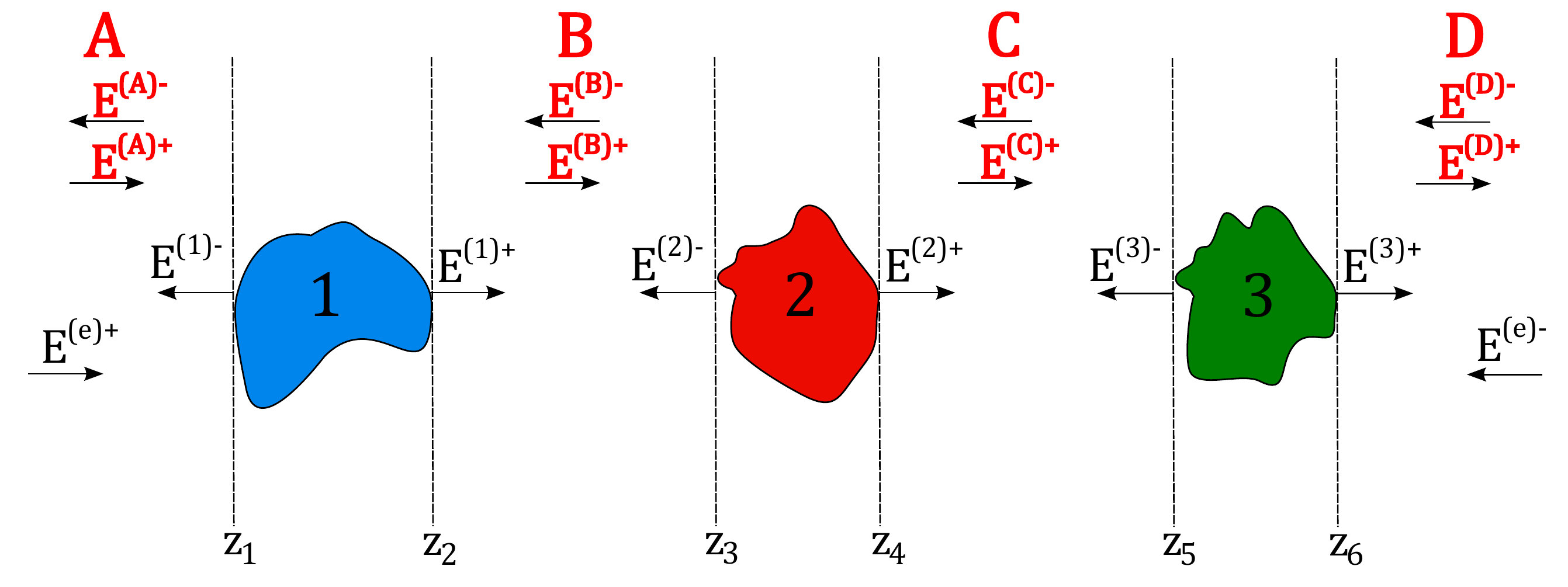}
\caption{Geometry of the three-body system. Bodies 1, 2 and 3 are respectively contained in the strips $z_1\leq z\leq z_2$, $z_3\leq z\leq z_4$ and $z_5\leq z\leq z_6$. This defines the four regions A, B, C
and D.}\label{Fig1}\end{figure}

The system we consider is made of three bodies, labeled with indexes 1, 2 and 3. The bodies have arbitrary geometries and material properties, as depicted in Fig. \ref{Fig1}. As we will see in the following, these properties will be accounted for by means of the classical electromagnetic reflection and transmission operators associated to each body. Moreover, we assume that each body $i$ is kept at a fixed temperature $T_i$ by external energy sources and that the three-body system is immersed in a environment characterized by a fourth (in general different) temperature $T_\text{e}$. Besides, we assume here that two parallel infinite planes can be found separating the couple of bodies (1,2) and (2,3). This assumption, verified in any typical experimental configuration, allows us to use a plane-wave basis and can be in principle relaxed by an appropriate change of basis.
In the plane-wave description we adopt from now on a single mode of the field is identified by the set of variables $(\omega,\mathbf{k},p,\phi)$, where $\omega$ is the frequency, $\mathbf{k}=(k_x,k_y)$ the component of the wavevector on the $xy$ plane (see Fig. \ref{Fig1}), $p$ the polarization index, taking the values $p=1,2$ corresponding to TE and TM modes respectively, and $\phi$ is the direction of propagation along the $z$ axis. In this approach, the $z$ component of the wavevector $k_z$ is a dependent variable, defined by
\begin{equation}k_z=\sqrt{\frac{\omega^2}{c^2}-\mathbf{k}^2},\end{equation}
while the complete wavevector $\mathbf{K}$ reads
\begin{equation}\mathbf{K}^\phi=(\mathbf{k},\phi k_z)=(k_x,k_y,\phi k_z).\end{equation}
For $k\leq\frac{\omega}{c}$, $k_z$ is real and the corresponding wave is propagative. On the contrary, for $k>\frac{\omega}{c}$, $k_z$ becomes imaginary and we have an evanescent wave: in this case $\phi$ is the direction along with the amplitude of the evanescent wave decays.

We now turn to the explicit expression of the electric fied, which we first decompose with respect to frequency, working only with positive frequencies
\begin{equation}\mathbf{E}(\mathbf{R},t)=2\Rea\Biggl[\int_0^{+\infty}\frac{d\omega}{2\pi}\exp(-i\omega t)\mathbf{E}(\mathbf{R},\omega)\Biggr].\end{equation}
The single-frequency component $\mathbf{E}(\mathbf{R},\omega)$ is then decomposed with respect to the parallel wavevector $\mathbf{k}$, the direction of propagation $\phi$ and the polarization $p$
\begin{equation}\label{DefE}\mathbf{E}(\mathbf{R},\omega)=\sum_{\phi,p}\int\frac{d^2\mathbf{k}}{(2\pi)^2}
\exp(i\mathbf{K}^\phi\cdot\mathbf{R})\hat{\bbm[\epsilon]}_p^\phi(\mathbf{k},\omega)E_p^\phi(\mathbf{k},\omega).\end{equation}
As a general rule, the sum on $\phi$ runs over the values
$\{+,-\}$, the sum on $p$ over the values $\{1,2\}$. For the polarization vectors
$\hat{\bbm[\epsilon]}_p^\phi(\mathbf{k},\omega)$ appearing in Eq. \eqref{DefE} we adopt the following standard definitions
\begin{equation}\label{PolVect}\begin{split}\hat{\bbm[\epsilon]}_\TE^\phi(\mathbf{k},\omega)&=\hat{\mathbf{z}}\times\hat{\mathbf{k}}=\frac{1}{k}(-k_y\hat{\mathbf{x}}+k_x\hat{\mathbf{y}})\\
\hat{\bbm[\epsilon]}_\TM^\phi(\mathbf{k},\omega)&=\frac{c}{\omega}\hat{\bbm[\epsilon]}_\TE^\phi(\mathbf{k},\omega)\times\mathbf{K}^\phi=\frac{c}{\omega}(-k\hat{\mathbf{z}}+\phi
k_z\hat{\mathbf{k}})\\\end{split}\end{equation} where
$\hat{\mathbf{x}}$, $\hat{\mathbf{y}}$ and $\hat{\mathbf{z}}$ are
the unit vectors along the directions $x$, $y$ and $z$
respectively and $\hat{\mathbf{k}}=\mathbf{k}/k$.

The expression of the single-frequency component of the magnetic field can be easily deduced from Maxwell's equations. It reads
\begin{equation}\label{DefB}\mathbf{B}(\mathbf{R},\omega)=\frac{1}{c}\sum_{\phi,p}\int\frac{d^2\mathbf{k}}{(2\pi)^2}
\exp(i\mathbf{K}^\phi\cdot\mathbf{R})\hat{\bbm[\beta]}_p^\phi(\mathbf{k},\omega)E_p^\phi(\mathbf{k},\omega)\end{equation}
where
\begin{equation}\hat{\bbm[\beta]}_p^\phi(\mathbf{k},\omega)=(-1)^p\hat{\bbm[\epsilon]}_{S(p)}^\phi(\mathbf{k},\omega)\end{equation}
being $S(p)$ the function which switches between the two
polarization, acting as $S(1)=2$ and $S(2)=1$.

\section{Maxwell stress tensor and Poynting vector}\label{SecT}

In order to calculate the Casimir-Lifshitz force and the heat transfer on each body we have to calculate the following surface integrals through a closed surface $\Sigma$ enclosing the body under scrutiny
\begin{equation}\label{FH}\begin{split}\mathbf{F}&=\int_\Sigma\langle\mathbb{T}(\mathbf{R},t)\rangle_\text{sym}\,\cdot d\bbm[\Sigma]\\
H&=-\int_{\Sigma}\langle\mathbf{S}(\mathbf{R},t)\rangle_\text{sym}\,\cdot
d\bbm[\Sigma]\end{split}\end{equation}
of the quantum symmetrized average of the Maxwell stress tensor $\mathbb{T}$ (having
cartesian components $T_{ij}$, with $i,j=x,y,z$) and the Poynting
vector $\mathbf{S}$. In classical electromagnetism, the definitions of these two quantities in SI units read
\begin{equation}\label{TijS}\begin{split}T_{ij}(\mathbf{R},t)&=\epsilon_0\Bigl[E_i(\mathbf{R},t)E_j(\mathbf{R},t)+c^2B_i(\mathbf{R},t)B_j(\mathbf{R},t)\\
&\,-\frac{1}{2}\Bigl(E^2(\mathbf{R},t)+c^2B^2(\mathbf{R},t)\Bigr)\delta_{ij}\Bigr],\\
\mathbf{S}(\mathbf{R},t)&=\epsilon_0c^2\mathbf{E}(\mathbf{R},t)\times\mathbf{B}(\mathbf{R},t).\end{split}\end{equation}
and the quantum symmetrized average value $\langle AB\rangle_\text{sym}$ is defined as
\begin{equation}\langle AB\rangle_\text{sym}=\frac{1}{2}\Bigl(\langle AB\rangle+\langle BA\rangle\Bigr)\end{equation}
being $\langle A\rangle$ an ordinary quantum average value. As shown in \cite{MessinaPRA11}, as a consequence of working in a plane-wave description, in order to calculate the $z$ component of the force and the heat transfer on a given body we only need the flux of $T_{zz}$ and $S_z$ through two planes $z=\bar{z}$ on the two sides of the body. We gather from now on the expressions relative to force and heat flux in a unique notation, by introducing an index $m$ whose value $m=1$ is associated to heat flux and $m=2$ to the force. In particular, we define
\begin{equation}\varphi_m(\bar{z})=\int_{z=\bar{z}}d^2\mathbf{r}\begin{cases}\langle S_z\rangle_\text{sym} & m=1\\\langle T_{zz}\rangle_\text{sym} & m=2\end{cases}
\end{equation}
Using the results of \cite{MessinaPRA11}, the \emph{generalized flux} can be cast under the form
\begin{equation}\label{GenFlux0}\begin{split}\varphi_m&(\bar{z})=-(-1)^m2\epsilon_0c^2\\
&\times\sum_p
\int\frac{d^2\mathbf{k}}{(2\pi)^2}\Biggl(\sum_{\phi=\phi'}\int_{ck}^{+\infty}\frac{d\omega}{2\pi}+\sum_{\phi\neq\phi'}\int_0^{ck}\frac{d\omega}{2\pi}\Biggr)\\
&\,\times\Bigl(\frac{\phi k_z}{\omega}\Bigr)^m\bra{p,\mathbf{k}}C^{\phi\phi'}\ket{p,\mathbf{k}}.\end{split}\end{equation}
In this expression we have introduced the matrix elements of the operator $C^{\phi\phi'}$, defined in terms of the correlation functions of the field amplitutes propagating in directions $\phi$ and $\phi'$ in the region where $\bar{z}$ is located throug the formula
\begin{equation}\label{CommC}\begin{split}\langle E^\phi_p&(\mathbf{k},\omega)E^{\phi'\dag}_{p'}(\mathbf{k}',\omega')\rangle_\text{sym}\\
&=\frac{1}{2}\langle E^\phi_p(\mathbf{k},\omega)E^{\phi'\dag}_{p'}(\mathbf{k}',\omega')+E^{\phi'\dag}_{p'}(\mathbf{k}',\omega')E^\phi_p(\mathbf{k},\omega)\rangle\\
&=2\pi\delta(\omega-\omega')\bra{p,\mathbf{k}}C^{\phi\phi'}\ket{p',\mathbf{k}'}.\end{split}\end{equation}
From \eqref{GenFlux0} we have
\begin{equation}\label{GenFlux}\begin{split}\varphi_m(\bar{z})&=-(-1)^m2\epsilon_0c^2\sum_{\phi\phi'}\Tr\Bigl\{\Bigl(\frac{\phi}{\omega}\Bigr)^mC^{\phi\phi'}\\
&\,\times\Bigl[\delta_{\phi\phi'}\mathcal{P}_m^\text{(pw)}+(1-\delta_{\phi\phi'})\mathcal{P}_m^\text{(ew)}\Bigr]\Bigr\},\end{split}\end{equation}
where $\delta_{\phi\phi'}$ is the Kronecker delta, we have introduced the trace operator
\begin{equation}\label{DefTrace}\Tr\mathcal{A}=\sum_p\int\frac{d^2\mathbf{k}}{(2\pi)^2}\int_0^{+\infty}\frac{d\omega}{2\pi}\bra{p,\mathbf{k}}\mathcal{A}\ket{p,\mathbf{k}}\end{equation}
and defined
\begin{equation}\bra{p,\mathbf{k}}\mathcal{P}_n^\text{(pw/ew)}\ket{p',\mathbf{k}'}=k_z^n\bra{p,\mathbf{k}}\Pi^\text{(pw/ew)}\ket{p',\mathbf{k}'}\end{equation}
$\Pi^\text{(pw)}$ ($\Pi^\text{(ew)}$) being the projector on the
propagative (evanescent) sector.

\section{Correlation functions of the total field}\label{SecCorr}

\subsection{Self-consistent scattering formulation}

In order to calculate the generalized flux \eqref{GenFlux} in any region we need the correlation functions of the total field. This field is a function of the source fields present in our system, namely the fields $E^{(i)\phi}$ emitted by the body $i$ and propagating in direction $\phi$ as well as the counterpropagating fields emitted by the environment $E^{\text{(e)}}\phi$, represented in Fig. \ref{Fig1}. The connection between total and source fields can be made explicit by introducing the reflection and transmission operators associated to each body and by writing down a self-consistent system of equations, in analogy with the method used in \cite{MessinaPRA11}, describing the multiple reflections occurring in the three-body configuration. For the reflection and transmission operators of body $i$ we will use the notations $\mathcal{R}^{(i)\phi}$ and $\mathcal{T}^{(i)\phi}$, where $\phi$ describes the direction of propagation of the outgoing field. For example, the matrix element
\begin{equation}\bra{p,\mathbf{k}}\mathcal{R}^{(i)\phi}\ket{p',\mathbf{k}'}\end{equation} gives the amplitude of the field mode $(\omega,\mathbf{k},p,\phi)$ reflected by body $i$ for an incoming field mode $(\omega,\mathbf{k}',p',-\phi)$: the frequency, implicitly contained in the reflection operator, is conserved since we consider only time-invariant configurations (for more details see \cite{MessinaPRA11}).

Using these operators, it is easy to write down the self-consistent system of equations giving the total field in each region
\begin{equation}\label{System}\begin{cases}E^{\text{(A)}+}=E^{\text{(e)}+}\\
E^{\text{(A)}-}=E^{(1)-}+\mathcal{R}^{(1)-}E^{\text{(e)}+}+\mathcal{T}^{(1)-}E^{\text{(B)}-}\\
E^{\text{(B)}+}=E^{(1)+}+\mathcal{R}^{(1)+}E^{\text{(B)}-}+\mathcal{T}^{(1)+}E^{\text{(e)}+}\\
E^{\text{(B)}-}=E^{(2)-}+\mathcal{R}^{(2)-}E^{\text{(B)}+}+\mathcal{T}^{(2)-}E^{\text{(C)}-}\\
E^{\text{(C)}+}=E^{(2)+}+\mathcal{R}^{(2)+}E^{\text{(C)}-}+\mathcal{T}^{(2)+}E^{\text{(B)}+}\\
E^{\text{(C)}-}=E^{(3)-}+\mathcal{R}^{(3)-}E^{\text{(C)}+}+\mathcal{T}^{(3)-}E^{\text{(e)}-}\\
E^{\text{(D)}+}=E^{(3)+}+\mathcal{R}^{(3)+}E^{\text{(e)}-}+\mathcal{T}^{(3)+}E^{\text{(C)}+}\\
E^{\text{(D)}-}=E^{\text{(e)}-}\end{cases}\end{equation}
We can now directly derive from Eq. \eqref{System} the total fields $E^{\text{(A)}\phi}$ and $E^{\text{(B)}\phi}$ in regions A and B. Due to the symmetry of the system, and using the invariance with respect to the exchange in the indexes of the fields and scattering operators
\begin{equation}(\text{A,B,C,D},+,-)\leftrightharpoons(\text{D,C,B,A},-,+),\end{equation}
we can derive total fields $E^{\text{(C)}\phi}$ and $E^{\text{(D)}\phi}$ in regions C and D.
This invariance is of course already manifest in the system of equations \eqref{System} itself.

\subsection{Many-body scattering operators}\label{SecMulti}

The solution of the system \eqref{System} can be conveniently provided in terms of many-body scattering operators taking into account the presence of two or three bodies at the same time. This was already done in the case of two bodies \cite{MessinaPRA11}, where the operators
\begin{equation}\begin{split}U^{(1,2)}=\sum_{n=0}^{+\infty}\bigl(\mathcal{R}^{(1)+}\mathcal{R}^{(2)-}\bigr)^n=(1-\mathcal{R}^{(1)+}\mathcal{R}^{(2)-})^{-1},\\
U^{(2,1)}=\sum_{n=0}^{+\infty}\bigl(\mathcal{R}^{(2)-}\mathcal{R}^{(1)+}\bigr)^n=(1-\mathcal{R}^{(2)-}\mathcal{R}^{(1)+})^{-1}.\end{split}\end{equation}
have been defined, describing the infinite series of multiple reflections in the \emph{cavity} formed by bodies 1 and 2. This interpretation explains why only $\mathcal{R}^{(1)+}$ and $\mathcal{R}^{(2)-}$ appear, i.e. the reflection operators of each body associated to the side on which the other body is located. We generalize here this definition by first introducing two-body reflection and transmission operators using the following intuitive definition given in the particular case of bodies 1 and 2
\begin{equation}\label{RT2}\begin{split}\mathcal{R}^{(12)+}&=\mathcal{R}^{(2)+}+\mathcal{T}^{(2)+}U^{(1,2)}\mathcal{R}^{(1)+}\mathcal{T}^{(2)-},\\
\mathcal{R}^{(12)-}&=\mathcal{R}^{(1)-}+\mathcal{T}^{(1)-}U^{(2,1)}\mathcal{R}^{(2)-}\mathcal{T}^{(1)+},\\
\mathcal{T}^{(12)+}&=\mathcal{T}^{(2)+}U^{(1,2)}\mathcal{T}^{(1)+},\\
\mathcal{T}^{(12)-}&=\mathcal{T}^{(1)-}U^{(2,1)}\mathcal{T}^{(2)-}.\end{split}\end{equation}
For example, the scattering processes participating to the definition of $\mathcal{R}^{(12)+}$ [reflection on the right-hand side of the couple of bodies (1,2)] are both single reflection on body 2 ($\mathcal{R}^{(2)+}$) and the transmission inside the cavity (1,2), the series of multiple reflections, and finally the transmission out of the cavity (described by the second term of the sum). An appropriate modification of Eq. \eqref{RT2} gives the definition of the two-body reflection and transmission operators for the couple (2,3). The definition \eqref{RT2} allows now to define iteratively three-body intracavity $U$ operators such as
\begin{equation}\begin{split}U^{(1,23)}=(1-\mathcal{R}^{(1)+}\mathcal{R}^{(23)-})^{-1},\\
U^{(23,1)}=(1-\mathcal{R}^{(23)-}\mathcal{R}^{(1)+})^{-1},\\
U^{(12,3)}=(1-\mathcal{R}^{(12)+}\mathcal{R}^{(3)-})^{-1},\\
U^{(3,12)}=(1-\mathcal{R}^{(3)-}\mathcal{R}^{(12)+})^{-1}.\end{split}\end{equation}
and finally three-body reflection and transmission operators
\begin{equation}\label{RT3}\begin{split}\mathcal{R}^{(123)+}&=\mathcal{R}^{(3)+}+\mathcal{T}^{(3)+}U^{(12,3)}\mathcal{R}^{(12)+}\mathcal{T}^{(3)-},\\
\mathcal{R}^{(123)-}&=\mathcal{R}^{(12)-}+\mathcal{T}^{(12)-}U^{(3,12)}\mathcal{R}^{(3)-}\mathcal{T}^{(12)+},\\
\mathcal{T}^{(123)+}&=\mathcal{T}^{(3)+}U^{(12,3)}\mathcal{T}^{(12)+},\\
\mathcal{T}^{(123)-}&=\mathcal{T}^{(12)-}U^{(3,12)}\mathcal{T}^{(3)-}.\end{split}\end{equation}
It is clear that in Eq. \eqref{RT3}, in order to write the three-body reflection and transmission operators, we have conceptually separated them in the two groups: (1,2) and (3). It is possible to show that the alternative choice [1 and (2,3)], giving for example
\begin{equation}\mathcal{R}^{(123)+}=\mathcal{R}^{(23)+}+\mathcal{T}^{(23)+}U^{(1,23)}\mathcal{R}^{(1)+}\mathcal{T}^{(23)-}.\end{equation}
leads to equivalent results. The choice of this subdivision is then only a matter of (theoretical or numerical) convenience.

The introduction of many-body reflection, transmission and intracavity operators has the advantage of allowing to write the total field in each region as a function of the source fields in a more compact and intuitive way.

\subsection{Total field in each region}\label{TotField}

Using the just defined many-body scattering operators and performing simple algebraic manipulations on Eq. \eqref{System} we obtain the total fields in regions A and B
\begin{equation}E^{\text{(A)}+}=E^{\text{(e)}+},\end{equation}
\begin{equation}\begin{split}E^{\text{(A)}-}&=\mathcal{T}^{(1)-}U^{(23,1)}\mathcal{R}^{(23)-}E^{(1)+}+E^{(1)-}\\
&\,+\mathcal{T}^{(12)-}U^{(3,12)}\mathcal{R}^{(3)-}E^{(2)+}+\mathcal{T}^{(1)-}U^{(23,1)}E^{(2)-}\\
&\,+\mathcal{T}^{(12)-}U^{(3,12)}E^{(3)-}+\mathcal{R}^{(123)-}E^{\text{(e)}+}\\
&\,+\mathcal{T}^{(123)-}E^{\text{(e)}-},\end{split}\end{equation}
\begin{equation}\begin{split}E^{\text{(B)}+}&=U^{(1,23)}\Bigl[E^{(1)+}+\mathcal{R}^{(1)+}\mathcal{T}^{(2)-}U^{(3,2)}\mathcal{R}^{(3)-}E^{(2)+}\\
&\,+\mathcal{R}^{(1)+}E^{(2)-}+\mathcal{R}^{(1)+}\mathcal{T}^{(2)-}U^{(3,2)}E^{(3)-}\\
&\,+\mathcal{T}^{(1)+}E^{\text{(e)}+}+\mathcal{R}^{(1)+}\mathcal{T}^{(23)-}E^{\text{(e)}-}\Bigr],\end{split}\end{equation}
\begin{equation}\begin{split}E^{\text{(B)}-}&=U^{(23,1)}\Bigl[\mathcal{R}^{(23)-}E^{(1)+}+\mathcal{T}^{(2)-}U^{(3,2)}\mathcal{R}^{(3)-}E^{(2)+}\\
&\,+E^{(2)-}+\mathcal{T}^{(2)-}U^{(3,2)}E^{(3)-}\\
&\,+\mathcal{R}^{(23)-}\mathcal{T}^{(1)+}E^{\text{(e)}+}+\mathcal{T}^{(23)-}E^{\text{(e)}-}\Bigr].\end{split}\end{equation}
This concludes the derivation of the total field in each region as a function of the source fields $E^{(i)\phi}$ ($i=1,2,3$, $\phi=+,-$) and the environmental field $E^{\text{(e)}\phi}$.

\subsection{Correlation functions of the source fields}\label{CorrSource}

In order to proceed further we now need to know the correlation function of the source fields. These are discussed and derived in \cite{MessinaPRA11}. In analogy with Eq. \eqref{CommC}, we introduce the two matrices
\begin{equation}\begin{split}\langle E^{(i)\phi}_p(\mathbf{k},\omega)&E^{(i)\phi'\dag}_{p'}(\mathbf{k}',\omega')\rangle_\text{sym}\\
&=2\pi\delta(\omega-\omega')\bra{p,\mathbf{k}}C^{(i)\phi\phi'}\ket{p',\mathbf{k}'},\\
\langle E^{\text{(e)}\phi}_p(\mathbf{k},\omega)&E^{\text{(e)}\phi'\dag}_{p'}(\mathbf{k}',\omega')\rangle_\text{sym}\\
&=2\pi\delta(\omega-\omega')\bra{p,\mathbf{k}}C^{\text{(e)}\phi\phi'}\ket{p',\mathbf{k}'}.\end{split}\end{equation}
We introduce the auxiliary function
\begin{equation}f_\alpha(\mathcal{R})=\begin{cases}\mathcal{P}_{-1}^{\text{(pw)}}-\mathcal{R}\mathcal{P}_{-1}^{\text{(pw)}}\mathcal{R}^\dag+\mathcal{R}\mathcal{P}_{-1}^{\text{(ew)}}-\mathcal{P}_{-1}^{\text{(ew)}}\mathcal{R}^\dag\\
 \hspace{5.3cm}\alpha=-1\\
\mathcal{P}_m^{\text{(pw)}}+(-1)^m\mathcal{R}^\dag\mathcal{P}_m^{\text{(pw)}}\mathcal{R}+\mathcal{R}^\dag\mathcal{P}_m^{\text{(ew)}}\\
+(-1)^m\mathcal{P}_m^{\text{(ew)}}\mathcal{R}\\
\hspace{4.3cm}\alpha=m\in\{1,2\}\end{cases}\end{equation}
and then express the correlation functions of the source fields under the form
\begin{equation}C^{(i)\phi\phi}=\frac{\omega}{2\varepsilon_0c^2}N_i\Bigl(f_{-1}(\mathcal{R}^{(i)\phi})-\mathcal{T}^{(i)\phi}\mathcal{P}_{-1}^{\text{(pw)}}\mathcal{T}^{(i)\phi\dag}\Bigr),\end{equation}
\begin{equation}\begin{split}C^{(i)\phi,-\phi}&=\frac{\omega}{2\varepsilon_0c^2}N_i\Bigl(-\mathcal{R}^{(i)\phi}\mathcal{P}_{-1}^{\text{(pw)}}\mathcal{T}^{(i)-\phi\dag}\\
&\,-\mathcal{T}^{(i)\phi}\mathcal{P}_{-1}^{\text{(pw)}}\mathcal{R}^{(i)-\phi\dag}+\mathcal{T}^{(i)\phi}\mathcal{P}_{-1}^{\text{(ew)}}\\
&\,-\mathcal{P}_{-1}^{\text{(ew)}}\mathcal{T}^{(i)-\phi\dag}\Bigr),\end{split}\end{equation}
\begin{equation}C^{\text{(e)}\phi\phi'}=\delta_{\phi\phi'}\frac{\omega}{2\varepsilon_0c^2}N_\text{e}\mathcal{P}_{-1}^{\text{(pw)}},\end{equation}
where for $\alpha\in\{1,2,3,4,\text{e}\}$ we have defined $N_\alpha=N(\omega,T_\alpha)$ and we have introduced the thermal population density
\begin{equation}N(\omega,T)=\frac{\hbar\omega}{2}\coth\Bigl(\frac{\hbar\omega}{2k_BT}\Bigr)=\hbar\omega\Bigl[\frac{1}{2}+n(\omega,T)\Bigr]\end{equation}
with
\begin{equation}n(\omega,T)=\frac{1}{e^{\frac{\hbar\omega}{k_BT}}-1}.\end{equation}

\subsection{Final result}

In Sec. \ref{TotField} we have derived the expression of the total field in each region as a function of the source fields. Being this relation always linear, we can write it under the general form
\begin{equation}E^{(\gamma)\phi}=\sum_{i=1}^3\sum_{\alpha=+,-}A^{(\gamma)\phi}_{i\alpha}E^{(i)\alpha}+\sum_{\alpha=+,-}B^{(\gamma)\phi}_\alpha E^{\text{(e)}\alpha},\end{equation}
where $\gamma\in\{\text{A,B,C,D}\}$. From this general expression we simply derive the general expression of the correlation functions of the total field in the region $\gamma$
\begin{equation}\begin{split}C_\gamma^{\phi\phi'}&=\sum_{i=1}^3\sum_{\alpha,\alpha'=+,-}A^{(\gamma)\phi}_{i\alpha}C^{(i)\alpha\alpha'}A^{(\gamma)\phi'\dag}_{i\alpha'}\\
&\,+\sum_{\alpha=+,-}B^{(\gamma)\phi}_\alpha C^{\text{(e)}}B^{(\gamma)\phi'\dag}_\alpha,\end{split}\end{equation}
where $C^{\text{(e)}}=C^{\text{(e)}\phi\phi}$. Being the coefficients $A^{(\gamma)\phi}_{i\alpha}$ and $B^{(\gamma)\phi}_\alpha$ of the decomposition known from Sec. \ref{TotField} and the source correlation functions known from Sec. \ref{CorrSource}, this concludes the derivation of the correlation functions of the total field in any region.

\section{Generalized flux in any region}\label{SecGenFlux}

Using the results obtained in the previous Section we are now able to derive the generalized flux $\varphi_m$ in the four regions A, B, C and D. In region A we have, after lenghty algebraic manipulations,
\begin{equation}\label{FluxA}\begin{split}&\varphi_m^{\text{(A)}}=-\Tr\Bigl\{\omega^{1-m}\\
&\Bigl[\Bigl(N_1+(-1)^mN_\text{e}\Bigr)\mathcal{P}_{m-1}^{\text{(pw)}}\\
&\,+N_{e1}\mathcal{R}^{(123)-}\mathcal{P}_{-1}^{\text{(pw)}}\mathcal{R}^{(123)-\dag}\mathcal{P}_m^{\text{(pw)}}\\
&\,+N_{e3}\mathcal{T}^{(123)-}\mathcal{P}_{-1}^{\text{(pw)}}\mathcal{T}^{(123)-\dag}\mathcal{P}_m^{\text{(pw)}}\\
&\,+N_{32}\mathcal{T}^{(12)-}U^{(3,12)}f_{-1}(\mathcal{R}^{(3)-})U^{(3,12)\dag}\mathcal{T}^{(12)-\dag}\mathcal{P}_m^{\text{(pw)}}\\
&\,+N_{21}\mathcal{T}^{(1)-}U^{(23,1)}f_{-1}(\mathcal{R}^{(23)-})U^{(23,1)\dag}\mathcal{T}^{(1)-\dag}\mathcal{P}_m^{\text{(pw)}}\Bigr]\Bigr\}\end{split}\end{equation}
where we have introduced the population differences $N_{\alpha\beta}=N_\alpha-N_\beta$, for $\alpha,\beta\in\{1,2,3,4,\text{e}\}$. This generalized flux is function of the four temperatures defined in the system, as well as of the geometrical and dielectric properties of the three bodies, embedded in their scattering operators. In region B we have
\begin{equation}\label{FluxB}\begin{split}&\varphi_m^{\text{(B)}}=-\Tr\Bigl\{\omega^{1-m}\\
&\,\Bigl[(-1)^mN_1U^{(1,23)}f_{-1}(\mathcal{R}^{(1)+})U^{(1,23)\dag}f_m(\mathcal{R}^{(23)-})\\
&\,+N_2U^{(23,1)}f_{-1}(\mathcal{R}^{(23)-})U^{(23,1)\dag}f_m(\mathcal{R}^{(1)+})\\
&\,+(-1)^mN_{e1}U^{(1,23)}\mathcal{T}^{(1)+}\mathcal{P}_{-1}^{\text{(pw)}}\mathcal{T}^{(1)+\dag}U^{(1,23)\dag}f_m(\mathcal{R}^{(23)-})\\
&\,+N_{e3}U^{(23,1)}\mathcal{T}^{(23)-}\mathcal{P}_{-1}^{\text{(pw)}}\mathcal{T}^{(23)-\dag}U^{(23,1)\dag}f_m(\mathcal{R}^{(1)+})\\
&\,+N_{32}U^{(23,1)}\mathcal{T}^{(2)-}U^{(3,2)}f_{-1}(\mathcal{R}^{(3)-})\\
&\,\times U^{(3,2)\dag}\mathcal{T}^{(2)-\dag}U^{(23,1)\dag}f_m(\mathcal{R}^{(1)+})\Bigr]\Bigr\}.\end{split}\end{equation}
For the two remaining fluxes (in regions C and D) we can once again exploit the symmetry of the system. These fluxes can be obtained from the results \eqref{FluxA} and \eqref{FluxB} in regions A and B with the index replacement
\begin{equation}(\text{A},\text{B},1,3,+,-)\longleftrightarrow(\text{D},\text{C},3,1,-,+)\end{equation}
and multiplying by $(-1)^m$.

\section{Thermal equilibrium}\label{SecThEq}

We will first consider the case of thermal equilibrium, assuming $T_1=T_2=T_3=T_\text{e}$. Indeed, even at equilibrium, the three-body force is quite unexplored, since on the a three-atom configuration has been investigate \cite{PassanteJPhysB98,RizzutoPRL07}.  All the fluxes of the Poynting vector are zero in each region
\begin{equation}\varphi_1^{(\gamma,\text{eq})}=0,\qquad\gamma=\text{A}, \text{B}, \text{C}, \text{D}.\end{equation}
This property implies the fact that, as physically evident, the heat flux on any body is zero when all the temperatures coincide. For the flux of the stress tensor we have
\begin{equation}\varphi_2^{\text{(A,eq)}}=-2\Tr\Bigl[N(\omega,T)\omega^{-1}\mathcal{P}_1^{\text{(pw)}}\Bigr]=\varphi_2^{\text{(D,eq)}},\end{equation}
\begin{equation}\begin{split}\varphi_2^{\text{(B,eq)}}&=-\Tr\Bigl\{\omega^{-1}N(\omega,T)\\
&\,\times\Bigl[U^{(1,23)}f_{-1}(\mathcal{R}^{(1)+})U^{(1,23)\dag}f_2(\mathcal{R}^{(23)-})\\
&+U^{(23,1)}f_{-1}(\mathcal{R}^{(23)-})U^{(23,1)\dag}f_2(\mathcal{R}^{(1)+})\Bigr]\Bigr\}\\
&=-2\Tr\Bigl[N(\omega,T)\omega^{-1}\mathcal{P}_1^{\text{(pw)}}\Bigr]\\
&\,-2\Rea\Tr\Bigl[k_z\omega^{-1}N(\omega,T)\Bigl(U^{(1,23)}\mathcal{R}^{(1)+}\mathcal{R}^{(23)-}\\
&\,+U^{(23,1)}\mathcal{R}^{(23)-}\mathcal{R}^{(1)+}\Bigr)\Bigr],\end{split}\end{equation}
\begin{equation}\begin{split}\varphi_2^{\text{(C,eq)}}&=-\Tr\Bigl\{\omega^{-1}N(\omega,T)\\
&\,\times\Bigl[U^{(3,12)}f_{-1}(\mathcal{R}^{(3)-})U^{(3,12)\dag}f_2(\mathcal{R}^{(12)+})\\
&+U^{(12,3)}f_{-1}(\mathcal{R}^{(12)+})U^{(12,3)\dag}f_2(\mathcal{R}^{(3)-})\Bigr]\Bigr\}\\
&=-2\Tr\Bigl[N(\omega,T)\omega^{-1}\mathcal{P}_1^{\text{(pw)}}\Bigr]\\
&\,-2\Rea\Tr\Bigl[k_z\omega^{-1}N(\omega,T)\Bigl(U^{(12,3)}\mathcal{R}^{(12)+}\mathcal{R}^{(3)-}\\
&\,+U^{(3,12)}\mathcal{R}^{(3)-}\mathcal{R}^{(12)+}\Bigr)\Bigr].\end{split}\end{equation}
We observe that the fluxes in the exterior regions A and D do not depend on the geometrical properties of the bodies, whereas this is not the case for the fluxes in the interior regions B and C.

\subsection{Force on body 1}

We are now ready to calculate the force acting on each body at thermal equilibrium, by taking the appropriate differences of fluxes calculated in the last Section. We remind here that for simplicity we deal only with the $z$ component of the force acting on each body. For the force acting on body 1 at thermal equilibrium we have, after manipulations analogous to the ones used in \cite{MessinaPRA11},
\begin{equation}\label{F1eq}\begin{split}F_{1z}^{\text{(eq)}}&=\varphi_2^{\text{(B,eq)}}-\varphi_2^{\text{(A,eq)}}\\
&=-2\Rea\Tr\Bigl[k_z\omega^{-1}N(\omega,T)\Bigl(U^{(1,23)}\mathcal{R}^{(1)+}\mathcal{R}^{(23)-}\\
&\,+U^{(23,1)}\mathcal{R}^{(23)-}\mathcal{R}^{(1)+}\Bigr)\Bigr]\end{split}\end{equation}
This result allows us to show explicitly the non-addivity of Casimir-Lifshitz forces even at thermal equilibrium. To this aim we can compare the force \eqref{F1eq} obtained in the present purely three-body approach to the sum of the two-body forces produced by each of the bodies 2 and 3 in absence of the other. These forces can be calculated using the results of \cite{MessinaPRA11} and their sum, i.e. the \emph{approximate additive result}, reads
\begin{equation}\begin{split}\tilde{F}_{1z}^{\text{(eq)}}&=-2\Rea\Tr\Bigl[k_z\omega^{-1}N(\omega,T)\\
&\,\times\Bigl(U^{(1,2)}\mathcal{R}^{(1)+}\mathcal{R}^{(2)-}+U^{(2,1)}\mathcal{R}^{(2)-}\mathcal{R}^{(1)+}\\
&\,+U^{(1,3)}\mathcal{R}^{(1)+}\mathcal{R}^{(3)-}+U^{(3,1)}\mathcal{R}^{(3)-}\mathcal{R}^{(1)+}\Bigr)\Bigr],\end{split}\end{equation}
which is manifestly different from the exact three-body result \eqref{F1eq}.

\subsection{Force on body 3}

The force acting on body 3 at thermal equilibrium can be deduced directly from the symmetry arguments discussed in the preceding sections. The exact force is given by
\begin{equation}\begin{split}F_{3z}^{\text{(eq)}}&=\varphi_2^{\text{(D,eq)}}-\varphi_2^{\text{(C,eq)}}\\
&=2\Rea\Tr\Bigl[k_z\omega^{-1}N(\omega,T)\Bigl(U^{(12,3)}\mathcal{R}^{(12)+}\mathcal{R}^{(3)-}\\
&\,+U^{(3,12)}\mathcal{R}^{(3)-}\mathcal{R}^{(12)+}\Bigr)\Bigr],\end{split}\end{equation}
while the sum of the forces produced by each of the bodies 1 and 2 in absence of the other would be
\begin{equation}\begin{split}\tilde{F}_{3z}^{\text{(eq)}}&=2\Rea\Tr\Bigl[k_z\omega^{-1}N(\omega,T)\\
&\,\times\Bigl(U^{(3,2)}\mathcal{R}^{(3)-}\mathcal{R}^{(2)+}+U^{(2,3)}\mathcal{R}^{(2)+}\mathcal{R}^{(3)-}\\
&\,+U^{(3,1)}\mathcal{R}^{(3)-}\mathcal{R}^{(1)+}+U^{(1,3)}\mathcal{R}^{(1)+}\mathcal{R}^{(3)-}\Bigr)\Bigr].\end{split}\end{equation}

\subsection{Force on body 2}

To conclude this Section, we finally give the force acting on body 2 at thermal equilibrium. We have
\begin{equation}\label{F2eq}\begin{split}F_{2z}^{\text{(eq)}}&=\varphi_2^{\text{(C,eq)}}-\varphi_2^{\text{(B,eq)}}\\
&=-2\Rea\Tr\Bigl[k_z\omega^{-1}N(\omega,T)\\
&\,\times\Bigl(U^{(12,3)}\mathcal{R}^{(12)+}\mathcal{R}^{(3)-}+U^{(3,12)}\mathcal{R}^{(3)-}\mathcal{R}^{(12)+}\\
&-U^{(1,23)}\mathcal{R}^{(1)+}\mathcal{R}^{(23)-}-U^{(23,1)}\mathcal{R}^{(23)-}\mathcal{R}^{(1)+}\Bigr)\Bigr].\end{split}\end{equation}
The sum of the two-body forces produced by bodies 1 and 3 reads
\begin{equation}\begin{split}\tilde{F}^{\text{(eq)}}_{2z}&=-2\Rea\Tr\Bigl[k_z\omega^{-1}N(\omega,T)\\
&\,\times\Bigl(U^{(2,3)}\mathcal{R}^{(2)+}\mathcal{R}^{(3)-}+U^{(3,2)}\mathcal{R}^{(3)-}\mathcal{R}^{(2)+}\\
&-U^{(1,2)}\mathcal{R}^{(1)+}\mathcal{R}^{(2)-}-U^{(2,1)}\mathcal{R}^{(2)-}\mathcal{R}^{(1)+}\Bigr)\Bigr].\end{split}\end{equation}
It is easy to verify that at thermal equilibrium the net force acting on the three-body system vanishes, i.e.
\begin{equation}F^{\text{(eq)}}_{2z}=-\Bigl(F^{\text{(eq)}}_{1z}+F^{\text{(eq)}}_{3z}\Bigr).\end{equation}

\section{Out of thermal equilibrium}\label{SecThNeq}

In Sec. \ref{SecGenFlux} we have derived the generalized flux in any region of the system. In particular, Eqs. \eqref{FluxA} and \eqref{FluxB} give the generalized fluxes in regions A and B, whereas the corresponding quantities in regions C and D can be easily obtained by symmetry arguments as discussed above. As discussed in Sec. \ref{SecT}, this knowledge allows us to deduce the force and the heat transfer on any body. In this context, one must remermber that the force is given by $\varphi_2$ on the right side of the body minus the same quantity on the left side, while the opposite difference of $\varphi_1$ provides the heat transfer.

\subsection{Force and heat transfer on body 1}\label{F1out}

In order to give force and heat transfer acting on body 1 we first recast the expression \eqref{FluxB} of the generalized flux in region B in the following form
\begin{equation}\label{FluxB2}\varphi_m^{\text{(B)}}=\delta_{m2}\Bigl[F_{1z}^{\text{(eq)}}(T_1)-2\Tr\Bigl(N_1\omega^{-1}\mathcal{P}_1^{\text{(pw)}}\Bigr)\Bigr]+\Delta_m^{\text{(B)}},\end{equation}
where
\begin{equation}\begin{split}&\Delta_m^{\text{(B)}}=-\hbar\Tr\Biggl\{\omega^{2-m}\\
&\times\Biggl[n_{21}U^{(23,1)}f_{-1}(\mathcal{R}^{(23)-})U^{(23,1)\dag}f_m(\mathcal{R}^{(1)+})\\
&\,+n_{32}U^{(23,1)}\mathcal{T}^{(2)-}U^{(3,2)}f_{-1}(\mathcal{R}^{(3)-})U^{(3,2)\dag}\mathcal{T}^{(2)-\dag}\\
&\times U^{(23,1)\dag}f_m(\mathcal{R}^{(1)+})\\
&\,+(-1)^mn_{e1}U^{(1,23)}\mathcal{T}^{(1)+}\mathcal{P}_{-1}^{\text{(pw)}}\mathcal{T}^{(1)+\dag}U^{(1,23)\dag}f_m(\mathcal{R}^{(23)-})\\
&\,+n_{e3}U^{(23,1)}\mathcal{T}^{(23)-}\mathcal{P}_{-1}^{\text{(pw)}}\mathcal{T}^{(23)-\dag}U^{(23,1)\dag}f_m(\mathcal{R}^{(1)+})\Biggr]\Biggr\}.\end{split}\end{equation}
The main advantage of the new expression \eqref{FluxB2} is that it separates an equilibrium contribution which is different from zero only for the force ($m=2$), and a non-equilibrium term $\Delta_m^{\text{(B)}}$ manifestly vanishing at thermal equilibrium.

The final result for both force and heat transfer on body 1 out of thermal equilibrium is the following
\begin{equation}\begin{split}H_1&=\Delta_{1,1}\\
F_{1z}&=F_{1z}^{\text{(eq)}}(T_1)+\Delta_{1,2}\end{split}\end{equation}
where the complete non-equilibrium contribution on body 1 is given by
\begin{equation}\begin{split}\Delta_{1,m}&=-2\delta_{m2}\Tr\Bigl(N(\omega,T_1)\omega^{-1}\mathcal{P}_1^{\text{(pw)}}\Bigr)\\
&\,+(-1)^m\Bigl(\Delta_m^{\text{(B)}}-\varphi_m^{\text{(A)}}\Bigr)\end{split}\end{equation}
which gives
\begin{widetext}
\begin{equation}\label{Delta1m}\begin{split}&\Delta_{1,m}=-(-1)^m\hbar\Tr\Biggl\{\omega^{2-m}\Biggl[n_{e1}\Bigl[U^{(23,1)}\mathcal{T}^{(23)-}\mathcal{P}_{-1}^{\text{(pw)}}\mathcal{T}^{(23)-\dag}U^{(23,1)\dag}\Bigl(f_m(\mathcal{R}^{(1)+})-\mathcal{T}^{(1)-\dag}\mathcal{P}_m^{\text{(pw)}}\mathcal{T}^{(1)-}\Bigr)\\
&+(-1)^m\Bigl(U^{(1,23)}\mathcal{T}^{(1)+}\mathcal{P}_{-1}^{\text{(pw)}}\mathcal{T}^{(1)+\dag}U^{(1,23)\dag}-\mathcal{P}_{-1}^{\text{(pw)}}\Bigr)f_m(\mathcal{R}^{(23)-})\\
&+\Bigl(\mathcal{R}^{(23)-}\mathcal{P}_{-1}^{\text{(pw)}}\mathcal{R}^{(23)-\dag}-\mathcal{R}^{(123)-}\mathcal{P}_{-1}^{\text{(pw)}}\mathcal{R}^{(123)-\dag}\Bigr)\mathcal{P}_m^{\text{(pw)}}\Bigr]\\
&+n_{21}U^{(23,1)}\Bigl(f_{-1}(\mathcal{R}^{(23)-})-\mathcal{T}^{(2)-}U^{(3,2)}f_{-1}(\mathcal{R}^{(3)-})U^{(3,2)\dag}\mathcal{T}^{(2)-\dag}\Bigr)U^{(23,1)\dag}\Bigl(f_m(\mathcal{R}^{(1)+})-\mathcal{T}^{(1)-\dag}\mathcal{P}_m^{\text{(pw)}}\mathcal{T}^{(1)-}\Bigr)\\
&+n_{31}U^{(23,1)}\mathcal{T}^{(2)-}U^{(3,2)}\Bigl(f_{-1}(\mathcal{R}^{(3)-})-\mathcal{T}^{(3)-}\mathcal{P}_{-1}^{\text{(pw)}}\mathcal{T}^{(3)-\dag}\Bigr)U^{(3,2)\dag}\mathcal{T}^{(2)-\dag}U^{(23,1)\dag}\Bigl(f_m(\mathcal{R}^{(1)+})-\mathcal{T}^{(1)-\dag}\mathcal{P}_m^{\text{(pw)}}\mathcal{T}^{(1)-}\Bigr)\Biggr]\Biggr\}.\end{split}\end{equation}
\end{widetext}
This expression is one of the main results of the paper. It gives the heat transfer ($m=1$) and the non-equilibrium contribution to the force ($m=2$) on body 1 for an arbitrary set of three bodies having arbitrary geometry, dielectric properties and temperatures and immersed in a thermal bath having a fourth different temperature. It is easy to verify that in the limit of absence of body 1 ($\mathcal{R}^{(1)\phi}=0$ and $\mathcal{T}^{(1)\phi}=1$) the non-equilibrium term $\Delta_{1,m}$ goes to zero, as well as the equilibrium force $F_{1z}^{\text{(eq)}}$ \red{of \ref{F1eq}}. Moreover, we observe that the contributions proportional to $n_{21}$ and $n_{31}$ go to zero in absence of body 2 and 3 respectively. In this sense they can be interpreted as the contribution to force and heat transfer associated to the exchange of energy and momentum between body 1 and each one of bodies 2 and 3. Nevertheless, one must keep in mind both that this subdivision is a matter of convenience and that the exchanges between body 1 and the two others are indeed dependent on the presence of the third body, as evident in Eq. \eqref{Delta1m}. This represents an explicit confirmation of the non-additive character of Casimir-Lifshitz force and radiative heat transfer.

We conclude this Section reminding that the force and heat transfer on body 3 out of thermal equilibrium can be obtained by simple symmetry arguments from the results we have just discussed concerning body 1.

\subsection{Force and heat transfer on body 2}

In analogy with what we have done in Sec. \ref{F1out} we first express the generalized flux in region B under the form
\begin{equation}\label{FluxB3}\varphi_m^{\text{(B)}}=\delta_{m2}\Bigl[F_{1z}^{\text{(eq)}}(T_2)-2\Tr\Bigl(N_2\omega^{-1}\mathcal{P}_1^{\text{(pw)}}\Bigr)\Bigr]+\Delta_m^{'\text{(B)}},\end{equation}
where
\begin{equation}\begin{split}&\Delta_m^{'\text{(B)}}=-\hbar\Tr\Biggl\{\omega^{2-m}\\
&\times\Biggl[(-1)^mn_{12}U^{(1,23)}f_{-1}(\mathcal{R}^{(1)+})U^{(1,23)\dag}f_m(\mathcal{R}^{(23)-})\\
&\,+n_{32}U^{(23,1)}\mathcal{T}^{(2)-}U^{(3,2)}f_{-1}(\mathcal{R}^{(3)-})U^{(3,2)\dag}\mathcal{T}^{(2)-\dag}\\
&\times U^{(23,1)\dag}f_m(\mathcal{R}^{(1)+})\\
&\,+(-1)^mn_{e1}U^{(1,23)}\mathcal{T}^{(1)+}\mathcal{P}_{-1}^{\text{(pw)}}\mathcal{T}^{(1)+\dag}U^{(1,23)\dag}f_m(\mathcal{R}^{(23)-})\\
&\,+n_{e3}U^{(23,1)}\mathcal{T}^{(23)-}\mathcal{P}_{-1}^{\text{(pw)}}\mathcal{T}^{(23)-\dag}U^{(23,1)\dag}f_m(\mathcal{R}^{(1)+})\Biggr]\Biggr\}.\end{split}\end{equation}

\begin{figure}[h!]
\includegraphics[height=5cm]{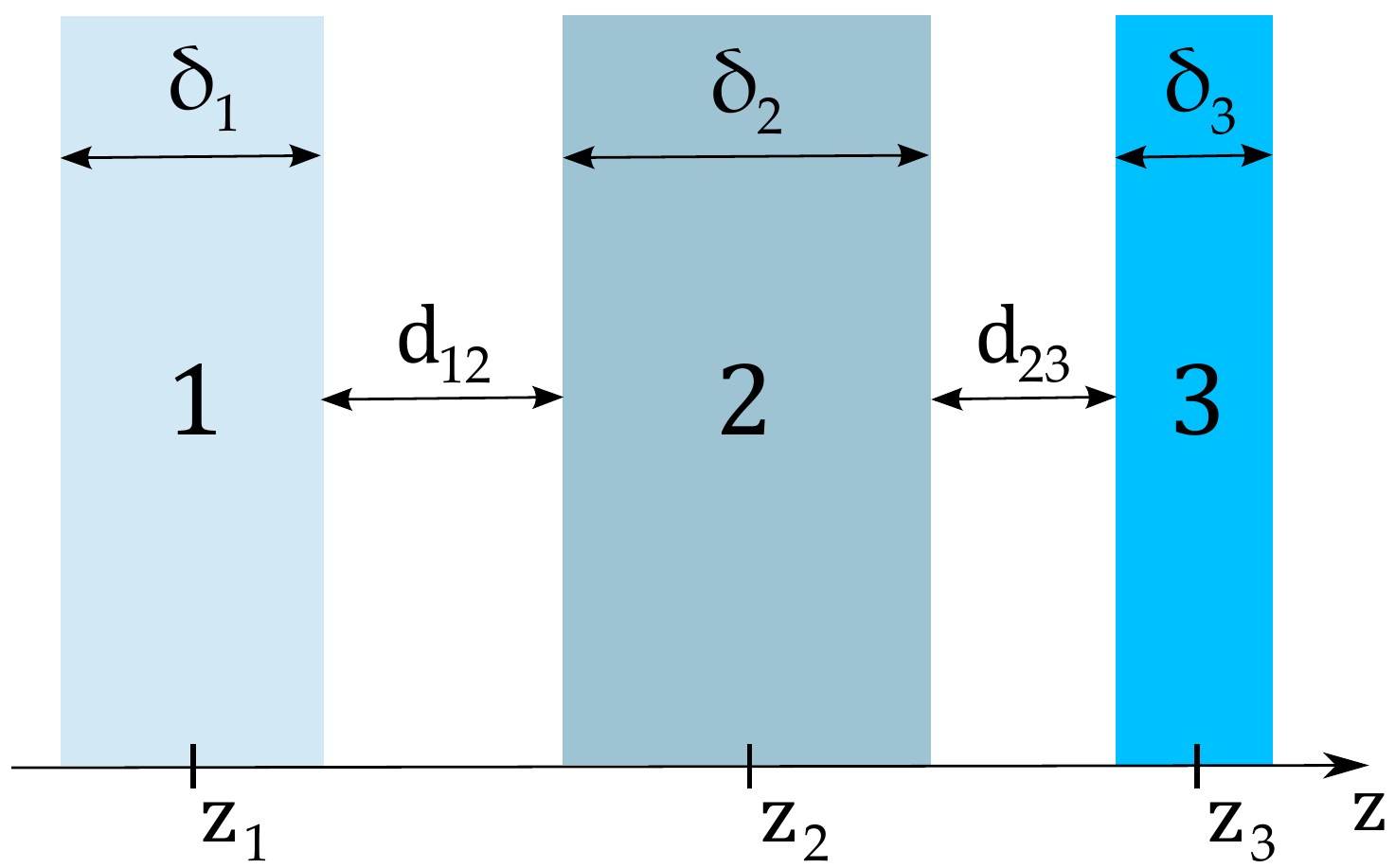}
\caption{Geometry of the three-slab configuration. The slab $i$ ($i=1,2,3$) is centered in $z_i$ and its thickness is $\delta_i$. The distance between slabs 1 and 2 is $d_{12}$, while the distance between slabs 2 and 3 is $d_{23}$.}\label{Fig3S}\end{figure}

This result allows us to write a similar expression for the generalized flux in region C, obtaining
\begin{equation}\label{FluxC}\varphi_m^{\text{(C)}}=\delta_{m2}\Bigl[F_{3z}^{\text{(eq)}}(T_2)-2\Tr\Bigl(N_2\omega^{-1}\mathcal{P}_1^{\text{(pw)}}\Bigr)\Bigr]+\Delta_m^{\text{(C)}},\end{equation}
where
\begin{equation}\begin{split}&\Delta_m^{\text{(C)}}=-\hbar\Tr\Biggl\{\omega^{2-m}\\
&\times\Biggl[(-1)^mn_{12}U^{(12,3)}\mathcal{T}^{(2)+}U^{(1,2)}f_{-1}(\mathcal{R}^{(1)+})U^{(1,2)\dag}\\
&\times \mathcal{T}^{(2)+\dag}U^{(12,3)\dag}f_m(\mathcal{R}^{(3)-})\\
&+n_{32}U^{(3,12)}f_{-1}(\mathcal{R}^{(3)-})U^{(3,12)\dag}f_m(\mathcal{R}^{(12)+})\\
&\,+(-1)^mn_{e1}U^{(12,3)}\mathcal{T}^{(12)+}\mathcal{P}_{-1}^{\text{(pw)}}\mathcal{T}^{(12)+\dag}\\
&\times U^{(12,3)\dag}f_m(\mathcal{R}^{(3)-})\\
&\,+n_{e3}U^{(3,12)}\mathcal{T}^{(3)-}\mathcal{P}_{-1}^{\text{(pw)}}\mathcal{T}^{(3)-\dag}U^{(3,12)\dag}f_m(\mathcal{R}^{(12)+})\Biggr]\Biggr\}.\end{split}\end{equation}
The final result for both force and heat transfer on body 2 out of thermal equilibrium is the following
\begin{equation}\begin{split}H_2&=\Delta_{2,1}\\
F_{2z}&=F_{2z}^{\text{(eq)}}(T_2)+\Delta_{2,2}\end{split}\end{equation}
where the non-equilibrium contribution is given by
\begin{equation}\Delta_{2,m}=(-1)^m\Bigl(\Delta_m^{\text{(C)}}-\Delta_m^{'\text{(B)}}\Bigr).\end{equation}

This gives
\begin{widetext}
\begin{equation}\label{Delta2m}\begin{split}&\Delta_{2,m}=-(-1)^m\hbar\Tr\Biggl\{\omega^{2-m}\\
&\times\Biggl[n_{e2}\Bigl[U^{(3,12)}\mathcal{T}^{(3)-}\mathcal{P}_{-1}^{\text{(pw)}}\mathcal{T}^{(3)-\dag}U^{(3,12)\dag}f_m(\mathcal{R}^{(12)+})-U^{(23,1)}\mathcal{T}^{(23)-}\mathcal{P}_{-1}^{\text{(pw)}}\mathcal{T}^{(23)-\dag}U^{(23,1)\dag}f_m(\mathcal{R}^{(1)+})\\
&+(-1)^mU^{(12,3)}\mathcal{T}^{(12)+}\mathcal{P}_{-1}^{\text{(pw)}}\mathcal{T}^{(12)+\dag}U^{(12,3)\dag}f_m(\mathcal{R}^{(3)-})-(-1)^mU^{(1,23)}\mathcal{T}^{(1)+}\mathcal{P}_{-1}^{\text{(pw)}}\mathcal{T}^{(1)+\dag}U^{(1,23)\dag}f_m(\mathcal{R}^{(23)-})\Bigr]\\
&+n_{12}(-1)^m\Bigl(f_{-1}(\mathcal{R}^{(1)+})-\mathcal{T}^{(1)+}\mathcal{P}_{-1}^{\text{(pw)}}\mathcal{T}^{(1)+\dag}\Bigr)\Bigl(U^{(1,2)\dag}\mathcal{T}^{(2)+\dag}U^{(12,3)\dag}f_m(\mathcal{R}^{(3)-})U^{(12,3)}\mathcal{T}^{(2)+}U^{(1,2)}\\
&-U^{(1,23)\dag}f_m(\mathcal{R}^{(23)-})U^{(1,23)}\Bigr)\\
&+n_{32}\Bigl(f_{-1}(\mathcal{R}^{(3)-})-\mathcal{T}^{(3)-}\mathcal{P}_{-1}^{\text{(pw)}}\mathcal{T}^{(3)-\dag})\Bigr)\Bigl(-U^{(3,2)\dag}\mathcal{T}^{(2)-\dag}U^{(23,1)\dag}f_m(\mathcal{R}^{(1)+})U^{(23,1)}\mathcal{T}^{(2)-}U^{(3,2)}\\
&+U^{(3,12)\dag}f_m(\mathcal{R}^{(12)+})U^{(3,12)}\Bigr)\Biggr]\Biggr\}\end{split}\end{equation}
\end{widetext}
We conclude by remarking that this expression goes to zero in absence of body 2 and, in analogy with what we have seen before, that the terms proportional to $n_{12}$ and $n_{32}$ go to zero in absence of bodies 1 and 3 respectively.

\section{Three parallel slabs}\label{Sec3s}

We now apply the general formulas for the force and heat transfer between three arbitrary bodies to the specific case of three parallel planar slabs of finite thickness. The slabs are identified with the indexes 1, 2 and 3 and their thicknesses are $\delta_1$, $\delta_2$ and $\delta_3$ respectively. The coordinates of the centers are $z_1$, $z_2$ and $z_3$ respectively, as represented in Fig. \ref{Fig3S}. As a consequence, the distances between adjacent slabs are given by
\begin{equation}\begin{split}d_{12}&=z_2-\frac{\delta_2}{2}-\Bigl(z_1+\frac{\delta_1}{2}\Bigr),\\
d_{23}&=z_3-\frac{\delta_3}{2}-\Bigl(z_2+\frac{\delta_2}{2}\Bigr).\end{split}\end{equation}

As a consequence of the translational invariance with respect to axes $x$ and $y$, the reflection and transmission operators of each slab are diagonal in the $(\mathbf{k},p)$ basis. Then, for a given frequency $\omega$ we have (see also \cite{MessinaPRA11} and \cite{BellomoPRA13})
\begin{equation}\label{SlabScatt}\begin{split}\bra{p,\mathbf{k}}\mathcal{R}^{(i)\phi}\ket{p',\mathbf{k}'}&=(2\pi)^2\delta(\mathbf{k}-\mathbf{k}')\delta_{pp'}\rho^{(i)\phi}_p(\mathbf{k},\omega),\\
\rho^{(i)\phi}_p(\mathbf{k},\omega)&=\rho^{(i)}_p(\mathbf{k},\omega)e^{-2i\phi k_z\bigl(z_i+\phi\frac{\delta_i}{2}\bigr)},\\
\bra{p,\mathbf{k}}\mathcal{T}^{(i)\phi}\ket{p',\mathbf{k}'}&=(2\pi)^2\delta(\mathbf{k}-\mathbf{k}')\delta_{pp'}\tau^{(i)}_p(\mathbf{k},\omega).\end{split}\end{equation}
We observe, as discussed more in detail in \cite{MessinaPRA11}, that the matrix element of $\mathcal{R}^{(i)\phi}$ depends on the side of the body we are considering, whereas the matrix element of $\mathcal{T}^{(i)\phi}$ is independent of $\phi$. The quantities $\rho^{(i)}_p$ and $\tau^{(i)}_p$ are defined in terms of the Fresnel reflection and transmission coefficients for a slab of the finite thickness $\delta_i$
\begin{equation}\begin{split}\rho^{(i)}_p(\mathbf{k},\omega)&=r^{(i)}_p(\mathbf{k},\omega)\frac{1-e^{2ik^{(i)}_z\delta_i}}{1-\bigl[r^{(i)}_p(\mathbf{k},\omega)\bigr]^2e^{2ik^{(i)}_z\delta_i}},\\
\tau^{(i)}_p(\mathbf{k},\omega)&=\frac{t^{(i)}_p(\mathbf{k},\omega)\bar{t}^{(i)}_p(\mathbf{k},\omega)e^{i(k^{(i)}_z-k_z)\delta_i}}{1-\bigl[r^{(i)}_p(\mathbf{k},\omega)\bigr]^2e^{2ik^{(i)}_z\delta_i}}.\\\end{split}\end{equation}
In these definitions we have introduced the $z$ component of the $\mathbf{K}$ vector inside the medium,
\begin{equation}k^{(i)}_z=\sqrt{\varepsilon_i(\omega)\frac{\omega^2}{c^2}-\mathbf{k}^2},\end{equation}
$\varepsilon_i(\omega)$ being the dielectric permittivity of the slab $i$, the ordinary vacuum-medium Fresnel reflection coefficients
\begin{equation}r^{(i)}_\text{TE}=\frac{k_z-k^{(i)}_z}{k_z+k^{(i)}_z},\qquad r^{(i)}_\text{TM}=\frac{\varepsilon_i(\omega)k_z-k^{(i)}_z}{\varepsilon_i(\omega)k_z+k^{(i)}_z},\end{equation}
as well as both the vacuum-medium (noted with $t$) and medium-vacuum (noted with $\bar{t}$) transmission coefficients
\begin{equation}\begin{split}t^{(i)}_\text{TE}&=\frac{2k_z}{k_z+k^{(i)}_z},\qquad\hspace{.3cm}t^{(i)}_\text{TM}=\frac{2\sqrt{\varepsilon_i(\omega)}k_{z}}{\varepsilon_i(\omega)k_z+k^{(i)}_z},\\
\bar{t}^{(i)}_\text{TE}&=\frac{2k^{(i)}_z}{k_z+k^{(i)}_z},\qquad\bar{t}^{(i)}_\text{TM}=\frac{2\sqrt{\varepsilon_i(\omega)}k^{(i)}_z}{\varepsilon_i(\omega)k_z+k^{(i)}_z}.\end{split}\end{equation}

The fact that the operators $\mathcal{R}^{(i)\phi}$ and $\mathcal{T}^{(i)\phi}$ are diagonal implies that the same property holds also for the many-body scattering operators as well as for the intracavity operators defined in Sec. \ref{SecMulti}. As a consequence, the relations defining these operator simply become scalar expressions connecting the matrix element $u$ of the operators $U$, and the matrix elements $\rho$ and $\tau$ of the reflection and transmission operators respectively. We start with the two-body intracavity operators, for which we have
\begin{equation}\begin{split}u^{(1,2)}&=u^{(2,1)}=\Bigl(1-\rho^{(1)}\rho^{(2)}e^{2ik_zd_{12}}\Bigr)^{-1},\\
u^{(2,3)}&=u^{(3,2)}=\Bigl(1-\rho^{(2)}\rho^{(3)}e^{2ik_zd_{23}}\Bigr)^{-1}.\end{split}\end{equation}
The two-body reflection operators for the couple (12) are given by
\begin{equation}\begin{split}\rho^{(12)+}&=\tilde{\rho}^{(12)+}e^{-2ik_z\bigl(z_2+\frac{\delta_2}{2}\bigr)}\\
\tilde{\rho}^{(12)+}&=\rho^{(2)}+\bigl(\tau^{(2)}\bigr)^2u^{(1,2)}\rho^{(1)}e^{2ik_z(d_{12}+\delta_2)},\\
\rho^{(12)-}&=\tilde{\rho}^{(12)-}e^{2ik_z\bigl(z_1-\frac{\delta_1}{2}\bigr)}\\
\tilde{\rho}^{(12)-}&=\rho^{(1)}+\bigl(\tau^{(1)}\bigr)^2u^{(1,2)}\rho^{(2)}e^{2ik_z(d_{12}+\delta_1)},\end{split}\end{equation}
while for the couple (23) we have
\begin{equation}\begin{split}\rho^{(23)+}&=\tilde{\rho}^{(23)+}e^{-2ik_z\bigl(z_3+\frac{\delta_3}{2}\bigr)}\\
\tilde{\rho}^{(23)+}&=\rho^{(3)}+\bigl(\tau^{(3)}\bigr)^2u^{(2,3)}\rho^{(2)}e^{2ik_z(d_{23}+\delta_3)},\\
\rho^{(23)-}&=\tilde{\rho}^{(23)-}e^{2ik_z\bigl(z_2-\frac{\delta_2}{2}\bigr)}\\
\tilde{\rho}^{(23)-}&=\rho^{(2)}+\bigl(\tau^{(2)}\bigr)^2u^{(2,3)}\rho^{(3)}e^{2ik_z(d_{23}+\delta_2)},\end{split}\end{equation}
The two-body transmission operators are given by
\begin{equation}\begin{split}\tau^{(12)}&=\tau^{(12)\phi}=\tau^{(1)}\tau^{(2)}u^{(1,2)},\\
\tau^{(23)}&=\tau^{(23)\phi}=\tau^{(2)}\tau^{(3)}u^{(2,3)}.\end{split}\end{equation}
We finally have the three-body intracavity operators
\begin{equation}\begin{split}u^{(1,23)}&=u^{(23,1)}=\Bigl(1-\rho^{(1)}\tilde{\rho}^{(23)-}e^{2ik_zd_{12}}\Bigr)^{-1},\\
u^{(12,3)}&=u^{(3,12)}=\Bigl(1-\tilde{\rho}^{(12)+}\rho^{(3)}e^{2ik_zd_{23}}\Bigr)^{-1},\end{split}\end{equation}
and the three-body reflection and transmission operators
\begin{equation}\begin{split}\rho^{(123)+}&=\tilde{\rho}^{(123)+}e^{-2ik_z\bigl(z_3+\frac{\delta_3}{2}\bigr)},\\
\tilde{\rho}^{(123)+}&=\tilde{\rho}^{(3)+}+\bigl(\tau^{(3)}\bigr)^2u^{(12,3)}\tilde{\rho}^{(12)+}e^{2ik_z(d_{23}+\delta_3)},\\
\rho^{(123)-}&=\tilde{\rho}^{(123)-}e^{2ik_z\bigl(z_1-\frac{\delta_1}{2}\bigr)},\\
\tilde{\rho}^{(123)-}&=\tilde{\rho}^{(1)-}+\bigl(\tau^{(1)}\bigr)^2u^{(1,23)}\tilde{\rho}^{(23)-}e^{2ik_z(d_{12}+\delta_1)},\\
\tau^{(123)}&=\tau^{(123)\phi}=\tau^{(12)}\tau^{(3)}u^{(12,3)}=\tau^{(1)}\tau^{(23)}u^{(1,23)}.\end{split}\end{equation}

\subsection{Pressure and heat transfer on slab 1}\label{PHTSlab1}

We are now ready to give the expression of the force and heat transfer acting on slab 1. As discussed in \cite{MessinaPRA11}, in presence of an infinite planar slab the result of the calculation is the pressure acting on the slab, the force being formally infinite. For the equilibrium three-body pressure we have from Eq. \ref{F1eq}:
\begin{equation}\begin{split}P_{1z}^{\text{(eq)}}&(T_1)=-\frac{1}{\pi^2}\Rea\sum_p\int_0^{+\infty} d\omega\frac{N(\omega,T_1)}{\omega}\\
&\times\int_0^{+\infty}dk\, k\,k_z\frac{\rho^{(1)}_p(\mathbf{k},\omega)\tilde{\rho}^{(23)-}_p(\mathbf{k},\omega)e^{2ik_zd_{12}}}{1-\rho^{(1)}_p(\mathbf{k},\omega)\tilde{\rho}^{(23)-}_p(\mathbf{k},\omega)e^{2ik_zd_{12}}}.\end{split}\end{equation}
This calculation can be performed, as usual, by means of the rotation to the imaginary axis.

It is useful to remember that the genuine two-body force acting on the slab 1 in presence of only slab 2 is:
\begin{equation}\begin{split}\tilde{P}_{1-2,z}^{\text{(eq)}}&(T_1)=-\frac{1}{\pi^2}\Rea\sum_p\int_0^{+\infty} d\omega\frac{N(\omega,T_1)}{\omega}\\
&\times\int_0^{+\infty}dk\, k\,k_z\frac{\rho^{(1)}_p(\mathbf{k},\omega)\rho^{(2)}_p(\mathbf{k},\omega)e^{2ik_zd_{12}}}{1-\rho^{(1)}_p(\mathbf{k},\omega)\rho^{(2)}_p(\mathbf{k},\omega)e^{2ik_zd_{12}}}.\end{split}\end{equation}
In presence of only slab 3, the pressure  $\tilde{P}_{1-3,z}^{\text{(eq)}}$ is obtained from the expression for $\tilde{P}_{1-2,z}^{\text{(eq)}}$ by modifying $\rho^{(2)}$ into $\rho^{(3)}$ and $d_{12}$ into $d_{12}+\delta_2+d_{23}$, i.e. the distance between slabs 1 and 3. It is also useful to define the sum
\begin{equation}\label{2bodytot}\tilde{P}_{1z}^{\text{(eq)}}=\tilde{P}_{1-2,z}^{\text{(eq)}}+\tilde{P}_{1-3,z}^{\text{(eq)}},\end{equation}
 which is the total \emph{two-body pressure} acting on body 1.

As for the non-equilibrium contribution, it can be written under the following form
\begin{equation}\label{Delta1m3s}\begin{split}&\Delta_{1,m}=-(-1)^m\frac{\hbar}{4\pi^2}\\
&\times\Bigl[A_{1,m}^{\text{(2,pw)}}(T_2)-A_{1,m}^{\text{(2,pw)}}(T_1)+A_{1,m}^{\text{(2,ew)}}(T_2)-A_{1,m}^{\text{(2,ew)}}(T_1)\\
&\,+A_{1,m}^{\text{(3,pw)}}(T_3)-A_{1,m}^{\text{(3,pw)}}(T_1)+A_{1,m}^{\text{(3,ew)}}(T_3)-A_{1,m}^{\text{(3,ew)}}(T_1)\\
&\,+A_{1,m}^{\text{(e,pw)}}(T_\text{e})-A_{1,m}^{\text{(e,pw)}}(T_1)\Bigr],\end{split}\end{equation}
where we have explicitly separated the contribution coming from evanescent waves (associated to the interactions between body 1 and the two others) and from propagative ones (including also the exchange with the environment). Using Eq. \eqref{Delta1m} we obtain the following explicit expressions of the functions appearing in Eq. \eqref{Delta1m3s} relative to the coupling with body 2
\begin{equation}\begin{split}&A_{1,m}^{\text{(2,pw)}}(T)=\sum_p\int_0^{+\infty} d\omega\,\omega^{2-m}n(\omega,T)\\
&\times\int_0^{\frac{\omega}{c}}dk\,k\,k_z^{m-1}|u^{(1,23)}|^2\Bigl(1+(-1)^m|\rho^{(1)}|^2-|\tau^{(1)}|^2\Bigr)\\
&\times\Bigl[1-|\tilde{\rho}^{(23)-}|^2-|\tau^{(2)}u^{(23)}|^2\Bigl(1-|\rho^{(3)}|^2\Bigr)\Bigr],\end{split}\end{equation}
\begin{equation}\begin{split}&A_{1,m}^{\text{(2,ew)}}(T)=2\text{i}^m\sum_p\int_0^{+\infty} d\omega\,\omega^{2-m}n(\omega,T)\\
&\times\int_{\frac{\omega}{c}}^{+\infty}dk\,k\,[\Ima(k_z)]^{m-1}|u^{(1,23)}|^2\Bigl(\rho^{(1)*}+(-1)^m\rho^{(1)}\Bigr)\\
&\times\Bigl[\Ima\bigl(\tilde{\rho}^{(23)-}\bigr)-|\tau^{(2)}u^{(23)}|^2\Ima\bigl(\rho^{(3)}\bigr)e^{-2\Ima(k_z)\bigl(d_{23}+\delta_2\bigr)}\Bigr]\\
&\times e^{-2\Ima(k_z)d_{12}},\end{split}\end{equation}
with body 3
\begin{equation}\begin{split}&A_{1,m}^{\text{(3,pw)}}(T)=\sum_p\int_0^{+\infty} d\omega\,\omega^{2-m}n(\omega,T)\\
&\times\int_0^{\frac{\omega}{c}}dk\,k\,k_z^{m-1}|u^{(1,23)}u^{(2,3)}\tau^{(2)}|^2\\
&\times\Bigl(1+(-1)^m|\rho^{(1)}|^2-|\tau^{(1)}|^2\Bigr)\Bigl(1-|\rho^{(3)}|^2-|\tau^{(3)}|^2\Bigr),\end{split}\end{equation}
\begin{equation}\begin{split}&A_{1,m}^{\text{(3,ew)}}(T)=2\text{i}^m\sum_p\int_0^{+\infty} d\omega\,\omega^{2-m}n(\omega,T)\\
&\times\int_{\frac{\omega}{c}}^{+\infty}dk\,k\,[\Ima(k_z)]^{m-1}|u^{(1,23)}u^{(2,3)}\tau^{(2)}|^2\\
&\times\Bigl(\rho^{(1)*}+(-1)^m\rho^{(1)}\Bigr)\Ima\bigl(\rho^{(3)}\bigr)e^{-2\Ima(k_z)\bigl(d_{12}+d_{23}+\delta_2\bigr)},\end{split}\end{equation}
and with the environment
\begin{equation}\begin{split}&A_{1,m}^{\text{(e,pw)}}(T)=\sum_p\int_0^{+\infty} d\omega\,\omega^{2-m}n(\omega,T)\\
&\times\int_0^{\frac{\omega}{c}}dk\,k\,k_z^{m-1}\\
&\times\Bigl[|u^{(1,23)}\tau^{(23)}|^2\Bigl(1+(-1)^m|\rho^{(1)}|^2-|\tau^{(1)}|^2\Bigr)\\
&+(-1)^m\Bigl(|u^{(1,23)}\tau^{(1)}|^2-1\Bigr)\Bigl(1+(-1)^m|\tilde{\rho}^{(23)-}|^2\Bigr)\\
&+|\tilde{\rho}^{(23)-}|^2-|\tilde{\rho}^{(123)-}|^2\Bigr].\end{split}\end{equation}

\begin{figure}[htb]
\includegraphics[height=7.5cm]{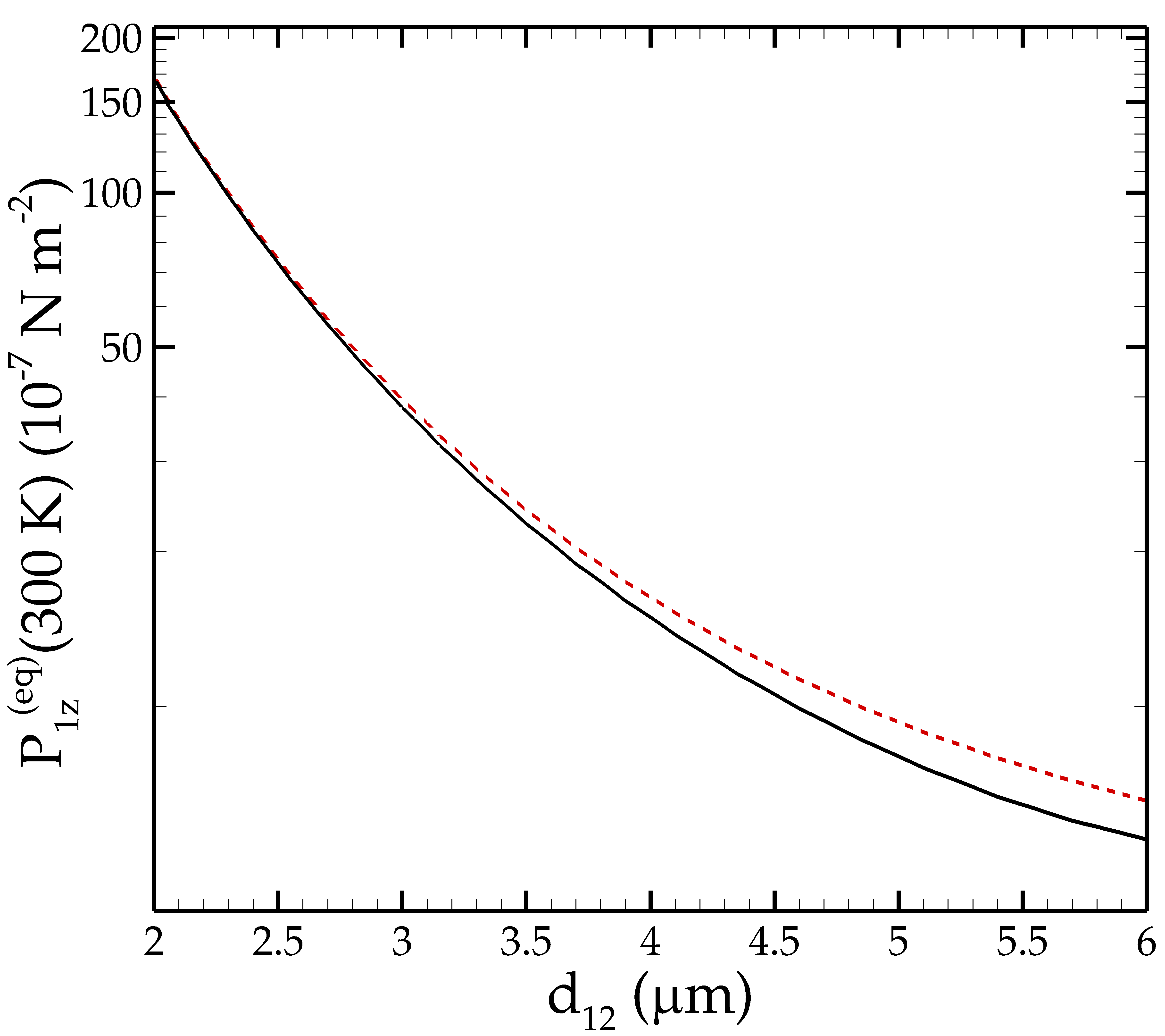}
\caption{Pressure acting on slab 1 at thermal equilibrium at $T=300\,$K. All slabs are made of sapphire. The black solid line is the three-body pressure $P_{1z}^{\text{(eq)}}$, while the red dashed line represents the sum of two-body contributions $\tilde{P}_{1z}^{\text{(eq)}}$ \ref{2bodytot}. The thickness of slab 2 is $\delta_2=1\,\mu\text{m}$, and the distance betweeen slabs 1 and 3 is fixed, so that we always have $d_{12}+\delta_2+d_{23}=7\,\mu\text{m}$ (see also Fig. \ref{FRelDiff}).}\label{FP1z}\end{figure}

\subsubsection{Numerical application: non-additivity of the force at thermal equilibrium}

\begin{figure*}[htb]
\includegraphics[height=11.5cm]{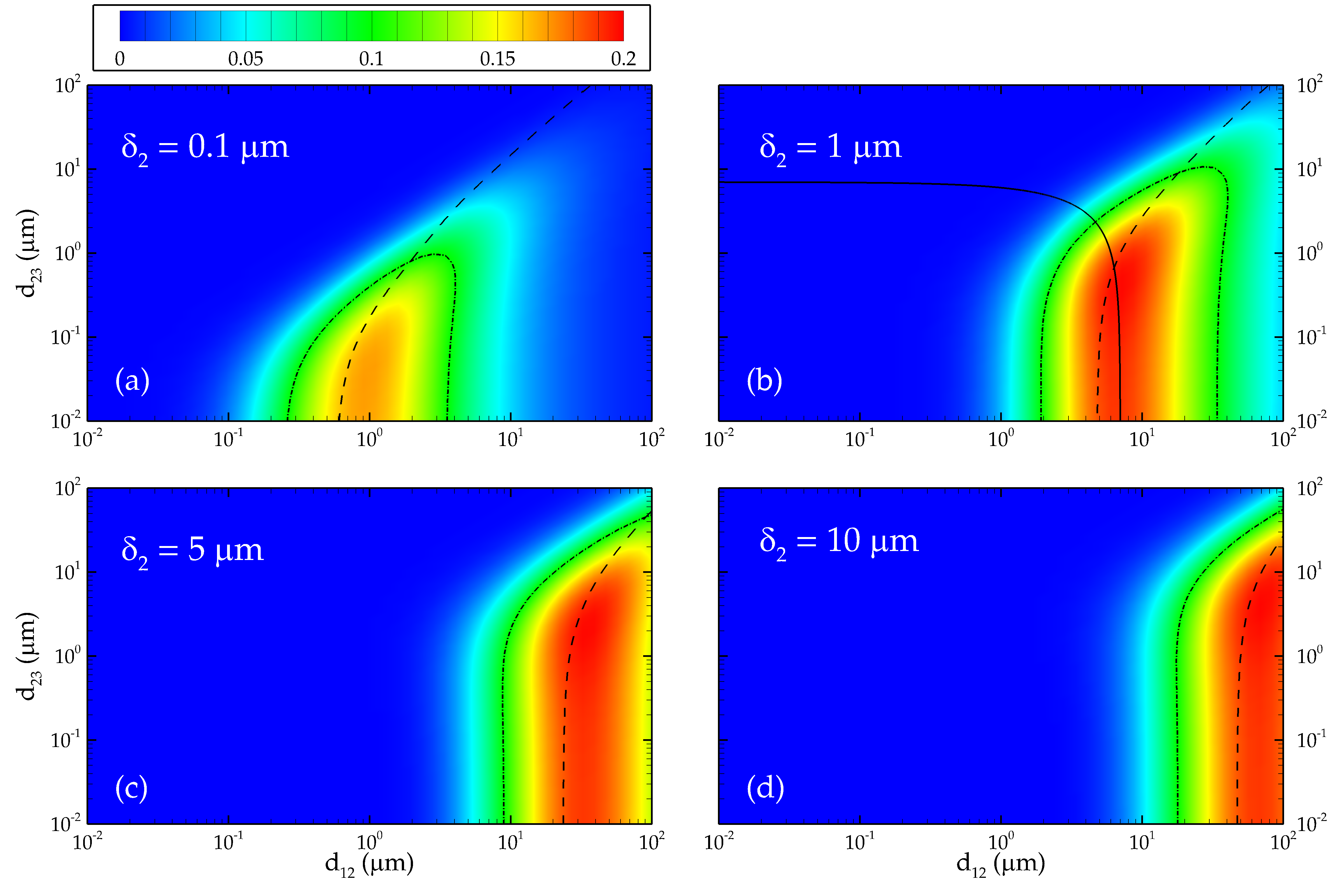}
\caption{Relative difference \eqref{DefRelDiff} of two- and three-body pressures acting on slab 1. The four panels (a), (b), (c) and (d) correspond to four different values of the thickness of slab 2, namely 100\,nm, 1\,$\mu$m, 5\,$\mu$m and 10\,$\mu$m respectively. In each panel the dot-dashed curve corresponds to a relative difference of 10\%, while the dashed line gives the point where the two-body coming from body 2 equals the one coming from body 3. Finally in (b) the solid line gives the points where $d_{12}+\delta_2+d_{23}=7\,\mu\text{m}$, used in Fig. \ref{FP1z}.}\label{FRelDiff}\end{figure*}

In this section, we will perform a quantitative analysis of the non-additivity of Casimir force, in the particular case of three parallel slabs. Here we will consider a configuration at thermal equilibrium at $T=300\,$K and three slabs made of sapphire (Al$_2$O$_3$). The numerical data used for this material are taken from \cite{Palik98}. Besides the values of the two- and three-body forces, we will consider the relative difference
\begin{equation}\label{DefRelDiff}\Delta P(d_{12},d_{23},\delta_2)=\frac{\tilde{P}_{1z}^{\text{(eq)}}-P_{1z}^{\text{(eq)}}}{P_{1z}^{\text{(eq)}}}\end{equation}
as a quantitative measure of non-additivity. Using \eqref{DefRelDiff}, we will be able to estimate in which regions of the parameters the additive approximation breaks down: in these regions, a fully three-body approach is mandatory in order to recover the correct value of the force.

We first plot in Fig. \ref{FP1z} the absolute values of the two- and three-body force acting on body 1. For this calculation, we have fixed the distance between slabs 1 and 3 to be 7\,$\mu$m and varied the distance $d_{12}$ between slabs 1 and 2, i.e. the position of slab 2 between the external ones. By representing the two forces for $d_{12}$ between 2 and 6 $\mu$m we see that a visible deviation is indeed present above 3 $\mu$m, and in particular the sum of two-body contributions always overestimates the exact result.

In order to get an quantitative insight on the non-additivity we have studied the relative difference \eqref{DefRelDiff} for four different values of the thickness $\delta_2$ and as a function of both $d_{12}$ and $d_{23}$. The results are shown in Fig. \ref{FRelDiff}. According to the values of the three parameters, relative differences up to 20\% can be observed. These plots allow to identify the regions of parameters where an additive approximation provides or not a good approximation for the total force acting on body 1. To this aim we have traced the dot-dashed line corresponding to a relative difference of 10\%. Moreover, we observe that the highest results are concentrated around the dashed curve associated to the points where $\tilde{P}_{1-2,z}^{\text{(eq)}}=\tilde{P}_{1-3,z}^{\text{(eq)}}$, i.e. where the two-body forces coming from bodies 2 and 3 give the same contribution.

\begin{figure}[htb]
\includegraphics[height=11.5cm]{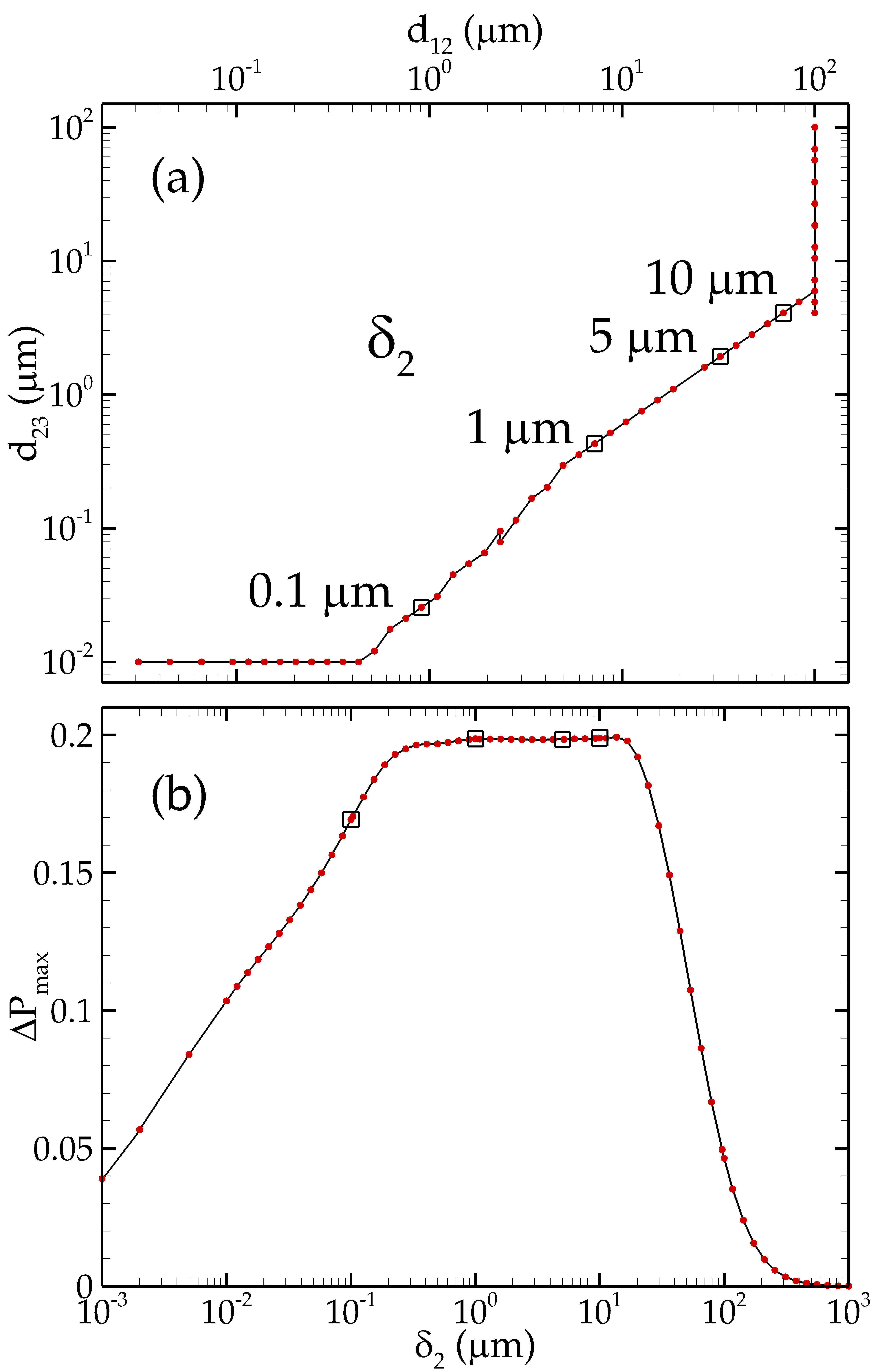}
\caption{Panel (a) describes the path followed as a function of $\delta_2$ by the couple $(d_{12},d_{23})$ realizing the highest value of the relative difference $\Delta P$, defined in Eq. \eqref{DefRelDiff}. Each red point of (a) corresponds to a red point of panel (b), from which the value of $\delta_2$ can be deduced. Panel (b) gives the highest possible value of $\Delta P$ for each value of $\delta_2$ in the region $[1\,\text{nm},1\,\text{mm}]$. The four squares in both panels correspond to the following values of $\delta_2$: 100\,nm, 1\,$\mu$m, 5\,$\mu$m and 10\,$\mu$m.}\label{FigDelta2}\end{figure}

We have finally considered the dependence of non-additivity on the thickness $\delta_2$. To this aim, for each value of $\delta_2$ between 1\,nm and 1\,mm we explore the regions $[10\,\text{nm},100\,\mu\text{m}]$ of $d_{12}$ and $d_{23}$ and find the couple $(d_{12},d_{23})$ corresponding to a maximum value of $\Delta P$. The path followed by this couple as a function of $\delta_2$ is represented in Fig. \ref{FigDelta2}(a). In this figure, the red points correspond to the points of Fig. \ref{FigDelta2}(b) (see caption for more details). We remark that for values of $\delta_2$ around 1\,$\mu$m both the values of $d_{12}$ and $d_{23}$ are of the order of some microns. Figure \ref{FigDelta2}(b) describes instead the highest values $\Delta P_{\text{max}}$ of the relative difference \eqref{DefRelDiff} as a function of $\delta_2$ (always considering $d_{12},d_{23}\in[10\,\text{nm},100\,\mu\text{m}]$). First of all we observe the presence of a plateau, roughly in the region $\delta_2\in[100\,\text{nm},10\,\mu\text{m}]$, where the relative difference can reach a significative value of 20\%. As remarked before, this region corresponds indeed to experimentally reasonable values of the couple of distances $(d_{12},d_{23})$. Finally, we see that both for large and small $\delta_2$ the relative difference goes to zero. This is physically expected, since both limiting cases correspond to a two-body interaction: for very large $\delta_2$ the force acting on body 1 is almost insensitive to the presence of body 3, whereas in the theoretical limit of $\delta_2$ going to 0 the force on body 1 is only due to body 3.

\subsection{Pressure and heat transfer on slab 2}

We now proceed dealing with pressure and heat transfer on the intermediate slab 2. We start with the equilibrium pressure, which reads
\begin{equation}\begin{split}P_{2z}^{\text{(eq)}}&(T_1)=-\frac{1}{\pi^2}\Rea\sum_p\int_0^{+\infty} d\omega\frac{N(\omega,T_1)}{\omega}\\
&\times\int_0^{+\infty}dk\, k\,k_z\Bigl[\frac{\rho^{(3)}_p(\mathbf{k},\omega)\tilde{\rho}^{(12)+}_p(\mathbf{k},\omega)e^{2ik_zd_{23}}}{1-\rho^{(3)}_p(\mathbf{k},\omega)\tilde{\rho}^{(12)+}_p(\mathbf{k},\omega)e^{2ik_zd_{23}}}\\
&-\frac{\rho^{(1)}_p(\mathbf{k},\omega)\tilde{\rho}^{(23)-}_p(\mathbf{k},\omega)e^{2ik_zd_{12}}}{1-\rho^{(1)}_p(\mathbf{k},\omega)\tilde{\rho}^{(23)-}_p(\mathbf{k},\omega)e^{2ik_zd_{12}}}\Bigr].\end{split}\end{equation}
The non-equilibrium contribution is again presented under the form of a sum of contributions coming from the bodies 1 and 3 and from the environment, as follows
\begin{equation}\label{Delta2m3s}\begin{split}&\Delta_{2,m}=-(-1)^m\frac{\hbar}{4\pi^2}\\
&\times\Bigl[A_{2,m}^{\text{(1,pw)}}(T_1)-A_{2,m}^{\text{(1,pw)}}(T_2)+A_{2,m}^{\text{(1,ew)}}(T_1)-A_{2,m}^{\text{(1,ew)}}(T_2)\\
&\,+A_{2,m}^{\text{(3,pw)}}(T_3)-A_{2,m}^{\text{(3,pw)}}(T_2)+A_{2,m}^{\text{(3,ew)}}(T_3)-A_{2,m}^{\text{(3,ew)}}(T_2)\\
&\,+A_{2,m}^{\text{(e,pw)}}(T_\text{e})-A_{2,m}^{\text{(e,pw)}}(T_2)\Bigr],\end{split}\end{equation}
where the individual terms are defines as
\begin{equation}\begin{split}&A_{2,m}^{\text{(1,pw)}}(T)=(-1)^m\sum_p\int_0^{+\infty} d\omega\,\omega^{2-m}n(\omega,T)\\
&\times\int_0^{\frac{\omega}{c}}dk\,k\,k_z^{m-1}\Bigl(1-|\rho^{(1)}|^2-|\tau^{(1)}|^2\Bigr)\\
&\times\Bigl[|\tau_2u^{(1,2)}u^{(12,3)}|^2\Bigl(1+(-1)^m|\rho^{(3)}|^2\Bigr)\\
&-|u^{(1,23)}|^2\Bigl(1+(-1)^m|\tilde{\rho}^{(23)-}|^2\Bigr)\Bigr],\end{split}\end{equation}
\begin{equation}\begin{split}&A_{2,m}^{\text{(1,ew)}}(T)=2\text{i}^m(-1)^m\sum_p\int_0^{+\infty} d\omega\,\omega^{2-m}n(\omega,T)\\
&\times\int_{\frac{\omega}{c}}^{+\infty}dk\,k\,[\Ima(k_z)]^{m-1}\Ima\bigl(\rho^{(1)}\bigr)\\
&\times\Bigl[-|u^{(1,23)}|^2\Bigl(\tilde{\rho}^{(23)-*}+(-1)^m\tilde{\rho}^{(23)-}\Bigr)\\
&+|\tau_2u^{(1,2)}u^{(12,3)}|^2\Bigl(\rho^{(3)*}+(-1)^m\rho^{(3)}\Bigr)e^{-2\Ima(k_z)\bigl(d_{23}+\delta_2\bigr)}\Bigr]\\
&\times e^{-2\Ima(k_z)d_{12}},\end{split}\end{equation}
\begin{equation}\begin{split}&A_{2,m}^{\text{(3,pw)}}(T)=\sum_p\int_0^{+\infty} d\omega\,\omega^{2-m}n(\omega,T)\\
&\times\int_0^{\frac{\omega}{c}}dk\,k\,k_z^{m-1}\Bigl(1-|\rho^{(3)}|^2-|\tau^{(3)}|^2\Bigr)\\
&\times\Bigl[-|\tau_2u^{(2,3)}u^{(1,23)}|^2\Bigl(1+(-1)^m|\rho^{(1)}|^2\Bigr)\\
&+|u^{(12,3)}|^2\Bigl(1+(-1)^m|\tilde{\rho}^{(12)+}|^2\Bigr)\Bigr],\end{split}\end{equation}
\begin{equation}\begin{split}&A_{2,m}^{\text{(3,ew)}}(T)=2\text{i}^m\sum_p\int_0^{+\infty} d\omega\,\omega^{2-m}n(\omega,T)\\
&\times\int_{\frac{\omega}{c}}^{+\infty}dk\,k\,[\Ima(k_z)]^{m-1}\Ima\bigl(\rho^{(3)}\bigr)\\
&\times\Bigl[|u^{(12,3)}|^2\Bigl(\tilde{\rho}^{(12)+*}+(-1)^m\tilde{\rho}^{(12)+}\Bigr)\\
&-|\tau_2u^{(2,3)}u^{(1,23)}|^2\Bigl(\rho^{(1)*}+(-1)^m\rho^{(1)}\Bigr)e^{-2\Ima(k_z)\bigl(d_{12}+\delta_2\bigr)}\Bigr]\\
&\times e^{-2\Ima(k_z)d_{23}},\end{split}\end{equation}
\begin{equation}\begin{split}&A_{2,m}^{\text{(e,pw)}}(T)=\sum_p\int_0^{+\infty} d\omega\,\omega^{2-m}n(\omega,T)\\
&\times\int_0^{\frac{\omega}{c}}dk\,k\,k_z^{m-1}\\
&\times\Bigl[|u^{(12,3)}\tau^{(3)}|^2\Bigl(1+(-1)^m|\tilde{\rho}^{(12)+}|^2\Bigr)\\
&-|u^{(1,23)}\tau^{(23)}|^2\Bigl(1+(-1)^m|\rho^{(1)}|^2\Bigr)\\
&-(-1)^m|u^{(1,23)}\tau^{(1)}|^2\Bigl(1+(-1)^m|\tilde{\rho}^{(23)-}|^2\Bigr)\\
&+(-1)^m|u^{(12,3)}\tau^{(12)}|^2\Bigl(1+(-1)^m|\rho^{(3)}|^2\Bigr)\Bigr].\end{split}\end{equation}

\subsubsection{Numerical application: equilibrium temperature of slab 2}

We will now present a numerical application of the formulas deduced in the previous section, focusing in particular on the heat transfer on the intermediate slab. In this context we will refer to an idea introduced in \cite{MessinaPRL12}, where we discussed how a three-body configuration can produce an enhancement of the energy transfer on one of the two external slab. In this work, in order to make the comparison of energy transfers between a two- and three-body configuration meaningful, we chose for the temperature of the intermediate slab (the one added with respect to the two-body case) the value making the flux on this intermediate slab zero. In this sense, the main point was that no additional external energy source (a thermostat) needed to be added to the system. In this section we will provide a study of how this equilibrium temperature varies as a function of the position of the intermediate slab for two given cavity width.

\begin{figure}[htb]
\includegraphics[height=7.5cm]{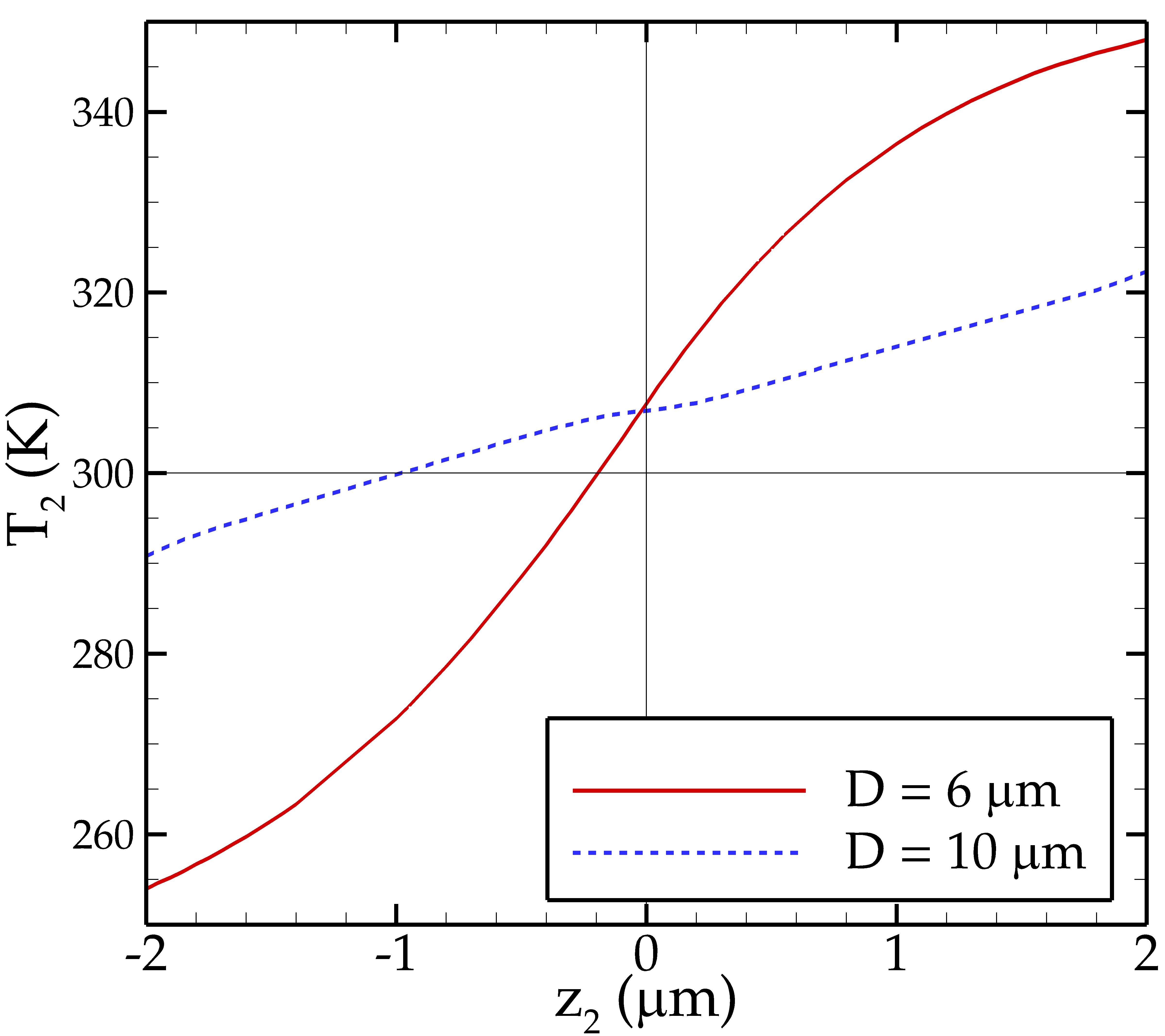}
\caption{Equilibrium temperature of slab 2 as a function of the its position $z_2$. The temperature of the external slabs are $T_1=250\,$K and $T_3=350\,$K, and two different values of the cavity width are shown: $D=6\,\mu$m (solid red line) and $D=10\,\mu$m (dashed blue line). All the three slabs are made of SiC and their thicknesses are $\delta_1=\delta_3=5\,\mu$m and $\delta_2=1\,\mu$m.}\label{FigT2eq}\end{figure}

The results are shown in Fig. \ref{FigT2eq}. We show the temperature at which the heat flux \eqref{Delta2m3s} (for $m=1$) vanishes (found using a simple bisection method) as a function of the position $z_2$ of slab 2 (see Fig. \ref{Fig3S}). For this calculation we have chosen three SiC slabs, whose dielectric permittivity is described using the simple model \cite{Palik98}
\begin{equation}\varepsilon(\omega)=\varepsilon_\infty\frac{\omega^2-\omega_\text{l}^2+i\Gamma\omega}{\omega^2-\omega_\text{t}^2+i\Gamma\omega},\end{equation}
where $\varepsilon_\infty=6.7$, $\omega_\text{l}=1.827\cdot10^{14}\,\mathrm{rad}\,\text{s}^{-1}$, $\omega_\text{t}=1.495\cdot10^{14}\,\mathrm{rad}\,\text{s}^{-1}$ and $\Gamma=0.9\cdot10^{12}\,\mathrm{rad}\,\text{s}^{-1}$. This model implies a surface phonon-polariton resonance at $\omega_p=1.787\cdot10^{14}\,\mathrm{rad}\,\text{s}^{-1}$. Finally, the thicknesses of the three slabs are $\delta_1=\delta_3=5\,\mu$m and $\delta_2=1\,\mu$m and we considered two different values of the cavity width, namely $D=6\,\mu$m and $D=10\,\mu$m.

In \cite{MessinaPRL12}, the scenario we considered was always symmetric, in the sense that the external slabs coincided and that slab 2 was always in the center of the cavity. Under this assumption we showed that in the spectrum of heat flux on body 2 the quantity $2n(\omega,T_2)-n(\omega,T_1)-n(\omega,T_3)$ factorizes. As a consequence, in the case of a quasi-monochromatic flux at frequency $\omega_0$, imposing that this quantity vanishes gives a first estimate of the equilibrium temperature. It is well known that in presence of surface resonance mode such as surface phonon polaritons the near-field heat transfer is quasi-monochromatic at this resonance frequency \cite{JoulainSurfSciRep05}. This enabled us to estimate the equilibrium temperature in \cite{MessinaPRL12} with a good precision. In the present case, this criterion gives an equilibrium temperature around 310\,K, representing a less precise estimate of the one, numerically found, near 307\,K. This can be explained by observing that in \cite{MessinaPRL12} the near-field assumption was much more verified, since the cavity width considered was at most $D=2.6\,\mu$m. Here, the difference between estimate and results proves that also non-resonant contributions are participating to the heat exchange.

One can note from the dependence of the equilibrium temperature on distance in  Fig. \ref{FigT2eq} the presence of a non-linear behavior, which reflects the non-linear behavior of $n(\omega,T)$. As expected, the equilibrium temperature tends to $T_1$ ($T_3$) when slab 2 approaches slab 1 (slab 3).

\section{Force on an atom between two slabs}\label{SecAt}

\begin{figure}[htb]
\includegraphics[height=5cm]{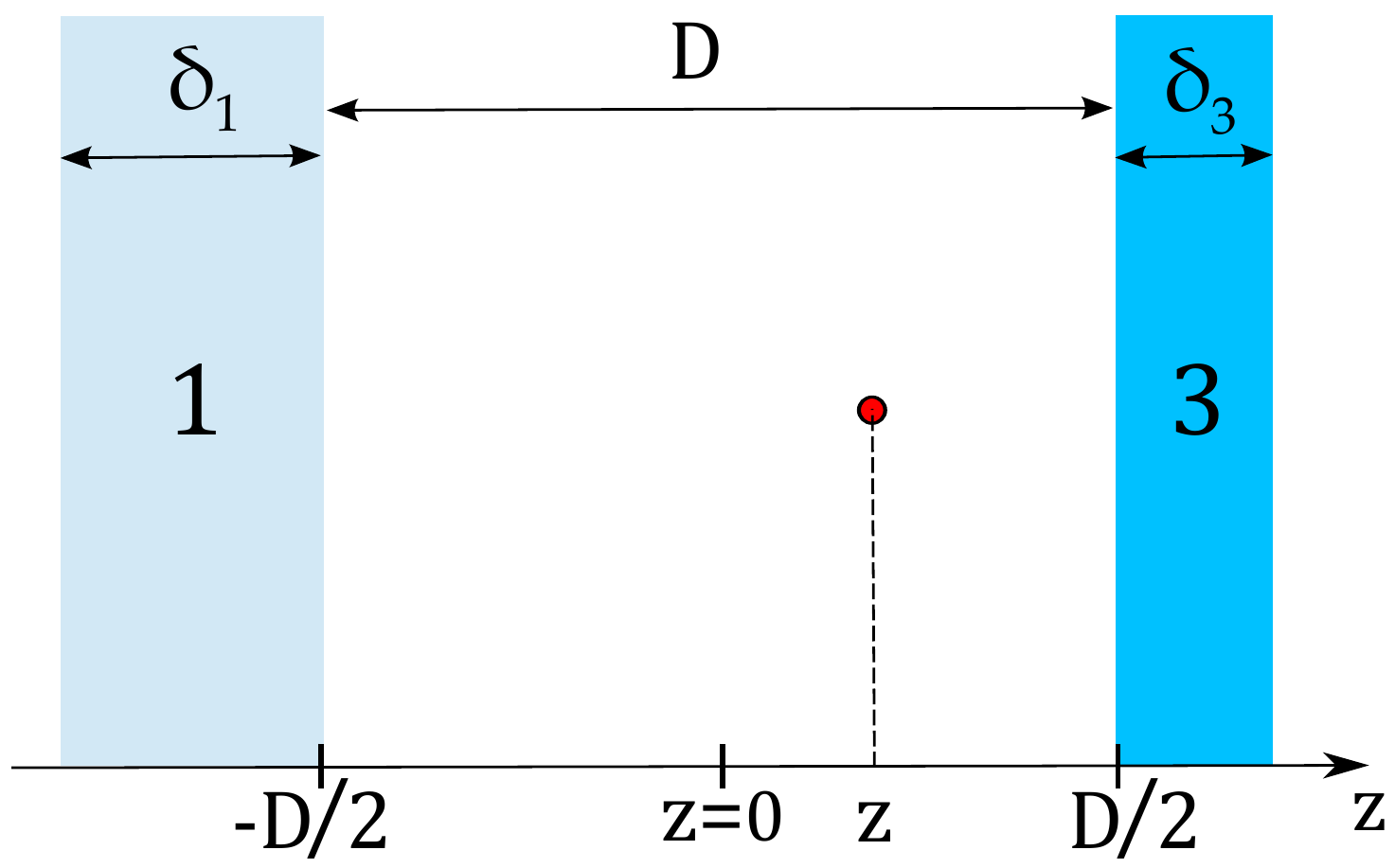}
\caption{Geometry of the configuration involving an atom between two slabs. The distance between slabs 1 and 3 is $D$, while $z$ is the atomic coordinate, $z=0$ being the plane in the middle of the cavity.}\label{FigAtSlabs}\end{figure}

In this section we will discuss the force acting on one atom placed between two parallel slabs in a configuration out of thermal equilibrium. The geometry of the system is described in Fig. \ref{FigAtSlabs}. We note with $D$ the width of the cavity, i.e. the distance between the two slabs. Moreover we note simply with $\mathbf{R}$ the atomic coordinate: in virtue of the cylindrical symmetry of our configuration, we expect all the results to depend only on the $z$ coordinate of the atom. Finally $z=0$ is the plane at the center of the cavity. Since the atom represents the body 2 in our formalism, for the sake of clarity we keep for the two slabs the indexes 1 and 3.

In order to deduce the force acting on the atom both at and out of thermal equilibrium we need the expression of its scattering (both reflection and transmission) operators. As discussed in \cite{MessinaPRA09}, these operators can be calculated within the dipole approximation by describing the atom as an induced dipole $\mathbf{d}(\omega)=\alpha(\omega)\mathbf{E}(\mathbf{R},\omega)$ proportional to the component of the electric field at frequency $\omega$ calculated at the atomic position $\mathbf{R}$, the proportionality factor being with the atomic dynamical polarizability $\alpha(\omega)$. It is convenient to write the atomic transmission operator under the form $\mathcal{T}_\text{A}^\phi=1+\tilde{\mathcal{T}}_\text{A}^\phi$ where we introduce the identity operator and an operator $\tilde{\mathcal{T}}_\text{A}^\phi$  which, in analogy with the reflection operator $\mathcal{R}_\text{A}^\phi$, gives no contribution in absence of the atom. Using these definition we have \cite{MessinaPRA09} \begin{equation}\label{AtScatt}\begin{split}\langle\mathbf{k},p|&\mathcal{R}_\text{A}^\phi(\omega)|\mathbf{k}',p'\rangle=\frac{i\omega^2\alpha(\omega)}{2\epsilon_0c^2k_z}\Bigl(\hat{\bbm[\epsilon]}_p^{\phi}(\mathbf{k},\omega)\cdot\hat{\bbm[\epsilon]}_{p'}^{-\phi}(\mathbf{k}',\omega)\Bigr)\\
&\,\times\exp[i(\mathbf{k}'-\mathbf{k})\cdot\mathbf{r}]\exp[-i\phi(k_z+k'_z)z],\\
\langle\mathbf{k},p|&\tilde{\mathcal{T}}_\text{A}^\phi(\omega)|\mathbf{k}',p'\rangle=\frac{i\omega^2\alpha(\omega)}{2\epsilon_0c^2k_z}\Bigl(\hat{\bbm[\epsilon]}_p^{\phi}(\mathbf{k},\omega)\cdot\hat{\bbm[\epsilon]}_{p'}^{\phi}(\mathbf{k}',\omega)\Bigr)\\
&\,\times\exp[i(\mathbf{k}'-\mathbf{k})\cdot\mathbf{r}]\exp[-i\phi(k_z-k'_z)z].\end{split}\end{equation}
Both operators are proportional to the dynamical polarizability $\alpha(\omega)$ and cancel in absence of the atom.

In order to calculate the force acting on the atom we have now to develop Eqs. \eqref{F2eq} and \eqref{Delta2m} using the slab and atomic scattering operators \eqref{SlabScatt} and \eqref{AtScatt} and keeping in mind that, coherently with the dipole approximation, we have to keep only the first order with respect to the atomic polarizability $\alpha(\omega)$. Using the fact that the scattering operators of both slabs are diagonal, it is easy to show that the equilibrium contribution simplifies to
\begin{equation}\begin{split}F_{2z}^{\text{(eq)}}=-4\Rea\Tr&\Bigl[k_z\omega^{-1}N(\omega,T)u^{(1,3)}\\
&\times\Bigl(\rho^{(3)-}\mathcal{R}_\text{A}^+-\rho^{(1)+}\mathcal{R}_\text{A}^-\Bigr)\Bigr].\end{split}\end{equation}
This expression equals the sum of the two two-body forces between the atom and each slab, apart from the presence of the intracavity reflection operator $U^{(1,3)}$. In this case, this operator clearly provides the origin of the non-additivity of Casimir forces, even in the case of a perturbative expansion. More explicitly, using the usual rotation to the imaginary axis, the previous expression explicitly becomes
\begin{widetext}
\begin{equation}\label{FAtEq}\begin{split}F_{2z}^{\text{(eq)}}&=\frac{k_\text{B}T\alpha(0)}{2\pi\epsilon_0}\int dk\,k^3u^{(1,3)}_\text{TM}(k,0)e^{-kD}\Bigl[\rho^{(3)}_\text{TM}(k,0)e^{2kz}-\rho^{(1)}_\text{TM}(k,0)e^{-2kz}\Bigr]\\
&\,+\frac{k_\text{B}T}{2\pi\epsilon_0c^2}\sum_{n=1}^{+\infty}\xi_n^2\alpha(i\xi_n)\int dk\,ke^{-\kappa_nD}\Bigl\{u^{(1,3)}_\text{TE}(k,i\xi_n)\Bigl[\rho^{(1)}_\text{TE}(k,i\xi_n)e^{-2\kappa_nz}-\rho^{(3)}_\text{TE}(k,i\xi_n)e^{2\kappa_nz}\Bigr]\\
&-u^{(1,3)}_\text{TM}(k,i\xi_n)\Bigl(\frac{2c^2k^2}{\xi_n^2}+1\Bigr)\Bigl[\rho^{(1)}_\text{TM}(k,i\xi_n)e^{-2\kappa_nz}-\rho^{(3)}_\text{TM}(k,i\xi_n)e^{2\kappa_nz}\Bigr]\Bigr\},\end{split}\end{equation}
where $\xi_n=2\pi k_\text{B}Tn/\hbar$ is the $n$-th Matsubara frequency and $\kappa_n=\sqrt{\xi_n^2/c^2+k^2}$. The non-equilibrium contribution reads
\begin{equation}\label{FAtNeq}\begin{split}&\Delta_{2,2}=-\frac{\hbar}{4\pi^2\epsilon_0c^2}\int d\omega\,\omega^2\sum_p\\
&\,\times\Biggl\{\Biggl[-n_{12}\int_0^{\frac{\omega}{c}}dk\,k|u^{(1,3)}|^2\Bigl(1-|\rho_1|^2-|\tau_1|^2\Bigr)\Bigl[2\Ima\bigl[\rho_3e^{2ik_z(\frac{D}{2}-z)}\bigr](\bbm[\varepsilon]^+\cdot\bbm[\varepsilon]^-)\Rea[\alpha(\omega)]+\Bigl(1-|\rho_3|^2\Bigr)\Ima[\alpha(\omega)]\Bigr]\\
&\,+2n_{12}\int_{\frac{\omega}{c}}^{+\infty}dk\,k|u^{(1,3)}|^2\Ima(\rho_1)\Bigl[\bigl(1-e^{-4\Ima(k_z)(\frac{D}{2}-z)}|\rho_3|^2\bigr)(\bbm[\varepsilon]^+\cdot\bbm[\varepsilon]^-)\Rea[\alpha(\omega)]e^{-2\Ima(k_z)(\frac{D}{2}+z)}\\
&\qquad\qquad-2\Ima(\rho_3)\Ima[\alpha(\omega)]e^{-2\Ima(k_z)D}\Bigr]\Biggr]-\Biggl[1\rightleftharpoons3,\frac{D}{2}\pm z\rightleftharpoons\frac{D}{2}\mp z\Biggr]\\
&\,+n_{e2}\int_0^{\frac{\omega}{c}}dk\,k|u^{(1,3)}|^2\Bigl[2\Bigl(|\tau_3|^2\Ima\bigl[\rho_1e^{2ik_z(\frac{D}{2}+z)}\bigr]-|\tau_1|^2\Ima\bigl[\rho_3e^{2ik_z(\frac{D}{2}-z)}\bigr]\Bigr)(\bbm[\varepsilon]^+\cdot\bbm[\varepsilon]^-)\Rea[\alpha(\omega)]\\
&\,-\Bigl(|\tau_1|^2\bigl(1-|\rho_3|^2\bigr)-|\tau_3|^2\bigl(1-|\rho_1|^2\bigr)\Bigr)\Ima[\alpha(\omega)]\Bigr]\Biggr\},\end{split}\end{equation}
\begin{center}\begin{figure}[htb]
\includegraphics[height=8.5cm]{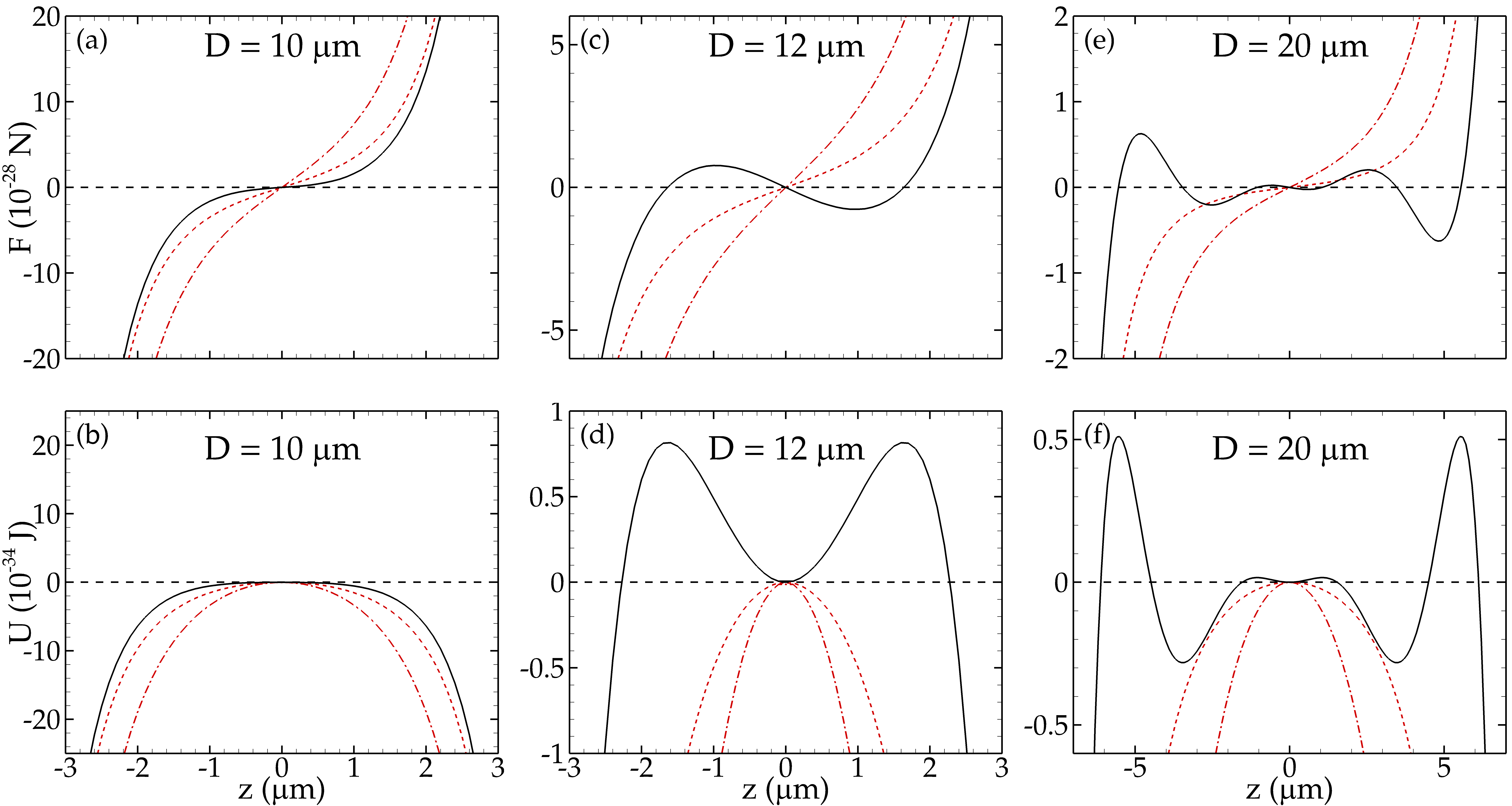}
\caption{Force [panels (a), (c) and (e)] and potential energy [panels (b), (d) and (f)] associated to a Rubidium atom in a cavity made of two sapphire slabs of equal thicknesses $\delta_1=\delta_3=5\,\mu$m. The temperatures are $T_1=T_3=300\,$K and $T_\text{e}=600\,$K. The figure shows three different values of the cavity width $D$. The presence of one or two potential minima are evident in panels (d) and (f), respectively.}\label{FigUF}\end{figure}\end{center}
\end{widetext}
where in the fourth line one has to subtract the first term in square brackets (the one associated to the difference $n_{12}$) after interchanging the indices 1 and 3 and $D/2\pm z$ with $D/2\mp z$. As a first example, we have used Eqs. \eqref{FAtEq} and \eqref{FAtNeq} to calculate the force acting on a Rubidium 87 atom \cite{BabbPRA04} and the associated potential energy in a symmetric configuration taking $\delta_1=\delta_3=5\,\mu$m and three different values of the cavity width $D$. Both slabs are made of sapphire, for which optical data are taken from \cite{Palik98}, the temperature of both slabs is $T_1=T_3=300\,$K, while the environmental temperature is $T_\text{e}=600\,$K. The results, presented in Fig. \ref{FigUF},  show the symmetry of our configuration, and in particular the fact that for any value of $D$ the force vanishes at $z=0$, i.e. in the center of the cavity (we remark that we have chosen $U(0)=0$). Moreover, we clearly see from panels (a), (c) and (e) that the forces changes its sign once for $D=10\,\mu$m, three times for $D=12\,\mu$m and five for $D=10\,\mu$m. Remarkably, this corresponds to the possibility of creating one or more potential wells for the atom by simply acting on the cavity width. Besides, this width also modulates the position and shape of the minima and maxima of the potential energy. The existence of these wells is an intuitive consequence of the repulsive character of the force already found and discussed in \cite{AntezzaPRL05} for an  atom in front of a single sapphire slab. In this paper the authors showed in detail how the combination of the environmental and slab radiation at different temperatures can produce a change in the sign of the force for distances of the order of $3\,\mu$m. It is the sum of two similar contributions (opposite in sign) that in the case of two slabs suggest the existence and tunability of potential minima and maxima. Nevertheless, the general case of two slabs having two different temperatures and material properties can be exactly solved only in the context of a purely three-body theory.

\begin{figure}[htb]
\includegraphics[height=7.5cm]{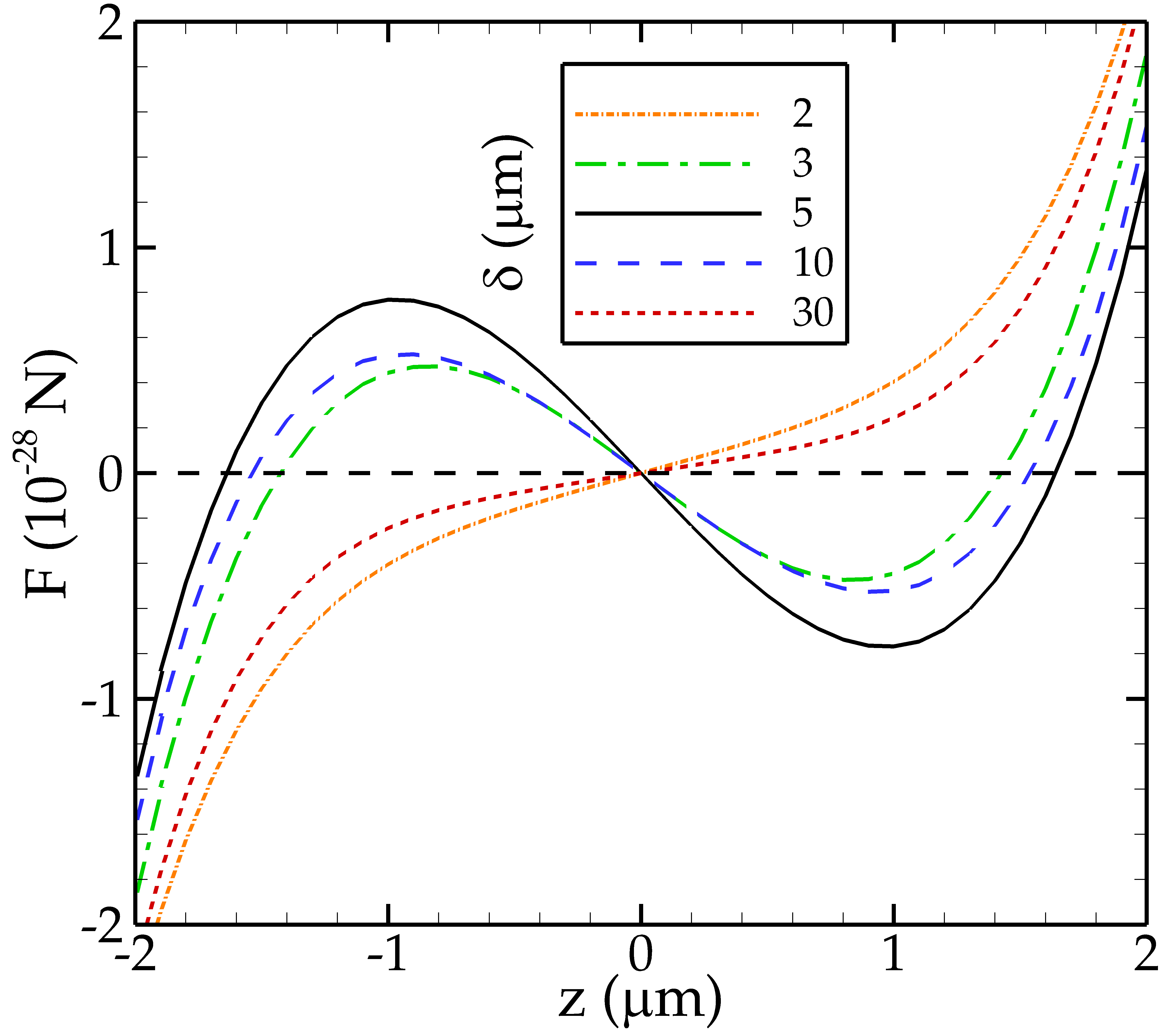}
\caption{Force acting on a Rubidium atom in a cavity of width $D=12\,\mu$m made of two sapphire slabs of equal thicknesses $\delta_1=\delta_3=\delta$. The temperatures are $T_1=T_3=300\,$K and $T_\text{e}=600\,$K. The figure shows five different values of the thickness $\delta$.}\label{FigAtDelta}\end{figure}

In order to get more insight into the possibilities offered by this scenario, we have studied the dependence of the force on the thickness $\delta_1=\delta_3=\delta$ of both slabs, as shown in Fig. \ref{FigAtDelta}. Considering five different values of $\delta$ in the range [2,30]$\,\mu$m we observe that the change in the sign of the force for $z<0$ and $z>0$ exists only for a limited range of thicknesses, and in particular disappears both for $\delta=2\,\mu$m and $\delta=30\,\mu$m. We remark here that for this configuration we have numerically verified that for all the possible parameters considered so far the additive result is almost undistinguishable for the exact three-body result at thermal equilibrium. This is not the case out of thermal equilibrium. For example, for $\delta=2\,\mu$m and $\delta=30\,\mu$m (see Fig. \ref{FigAtDelta}) the additive result predicts the three sign inversions of the force observed for $\delta=5\,\mu$m, whereas the exact result shows only one inversion, and as a consequence no potential well.

\begin{figure}[htb]
\includegraphics[height=7.5cm]{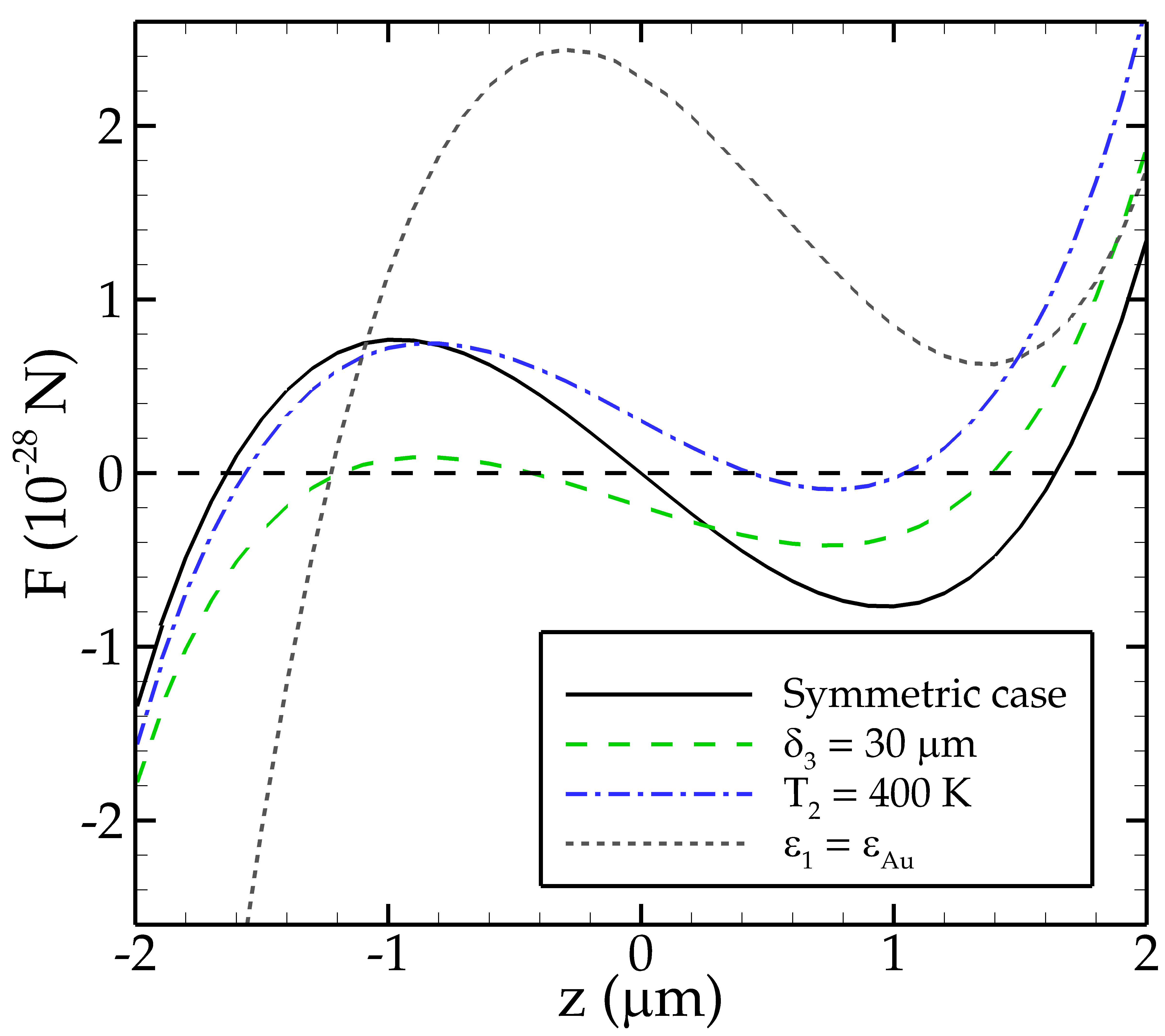}
\caption{Force acting on a rubidium atom in several different asymmetric configurations. The solid black line corresponds to the reference symmetric case (cavity width of $D=12\,\mu$m, sapphire walls having thicknesses $\delta_1=\delta+3=5\,\mu$m), the green long-dashed line is associated to $\delta_3=30\,\mu$m, the blue dot-dashed line to $T_2=400\,$K, the short-dashed grey line to the choice of gold for slab 1.}\label{FigAtAsymmetry}\end{figure}

We have finally investigated the possibility of controlling the shape of the potential well by producing on purpose an asymmetry in the geometrical, material or thermal configuration of the system. To this aim, we have chosen as a reference condition the case of a rubidium atom in a cavity having width $D=12\,\mu$m and made of sapphire walls having thicknesses $\delta_1=\delta+3=5\,\mu$m. With respect to this configuration, we have independently modified the thickness of the slab on the right ($\delta_3=30\,\mu$m), its temperature ($T_2=400\,$K) or replaced the material of slab 1 with gold. The results, shown in Fig. \ref{FigAtAsymmetry}, show that acting on geometrical, material or thermal properties is a promising tool to design the shape of the wells, their position and even to control their existence (for example for the parameters chosen here the presence of gold makes the well disappear).

From Fig. \ref{FigUF} we see that the depth of the potential wells is of the order of $10^{-34}\,$J, then far from being sufficient for an atomic trap for ground-state atoms, since the average kinetic energy $k_\text{B}T$ is of the order of $10^{-26}\,$J for temperatures in the range of mK. Since the depth of the potential wells is proportional to the atomic polarizability, the situation may completely change for Rydberg atoms.  For such kind of atoms, the fact of having a high value of the principal quantum number $n$ drastically modifies the value of the polarizability, scaling approximately as $n^7$ with respect to the ground-state one. As a consequence, even for relatively low values of $n$, for example of the order of 20, $\alpha$ increases by a factor around $10^{9}$, sufficient to realize an atomic trap. It is worth stressing that the huge increase in the size of these atoms may need to take into account also the quadrupole and octupole contributions to the atom-field interaction \cite{CrossePRA09,CrossePRA10}. We also remark that the trapping scheme we propose here is robust and universal, since uniquely based on thermal non-equilibrium configuration. It differs from the cavity trapping scheme discussed in \cite{EllingsenPRA09}, based on a resonant process exploiting molecular transitions.

\section{Conclusions}\label{SecConcl}

We have calculated the Casimir force in and out of thermal equilibrium and the radiative heat transfer in a system made of three bodies of arbitrary geometry, temperature and dielectric properties, immersed in an environment having a fourth (in general different) temperatures. To this aim, we have described each body through its classical reflection and transmission operators and deduced for each body a general analytic unified expression of force and heat transfer as a function of these individual operators. One of the main  advantage of this approach is that in order to calculate force and heat transfer one only needs to know the scattering operators of each body, which result indeed from individual single-body electromagnetic problems.

The exact solution of a three-body problem raises several interesting questions, the first of which is the range of validity of a simplified additive approach based on the knowledge of two-body results. This topic is discussed in our first numerical application, i.e. the calculation of the equilibrium force acting on an external slab of a system made of three parallel slabs. For this system we have shown that there exist indeed regions of the parameters where the additive result is a very good approximation, but this approach fails in some regions with an error as high as 20\%. As a consequence, it is in general relevant to know the exact result for the sake of experiment-theory comparison.

For the same system (three parallel slabs) we have also considered the heat transfer on the intermediate one for three given temperatures of the external slabs and of the environment. This has allowed us to calculate the equilibrium temperature of the intermediate slab at any position, i.e. the temperature which remains constant even in absence of an external source of energy. In particular, we have shown and briefly discussed that the temperature even in a symmetric configuration differs from the average of the temperatures of the external slabs.

Finally, we have considered the case of an atom in a planar cavity. For this system we have shown that the manipulation of the temperatures of the external slabs, of their dielectric properties as well as the lengths involved is able to modify qualitatively the shape of the force acting on the atom. Remarkably, for certain values of the parameters, it is possible to produce one or more potential wells. While for typical ground-state atoms the depth of the well is too small to produce a trap, this is not the case for Rydberg atoms.  In this case the combination of thermal non-equilibrium and of a the three-body configuration may realize a single or multiple atomic trap, with promising possibilities in terms of control of the depth and shape of the trap. Rydberg atoms are already manipulated in optical cavity for fundamental studies in quantum physics and for quantum information purposes \cite{GleyzesNature07,GuerlinNature07} with microwave cavities having sizes of the order of several centimeters. Here we deal with a trap and a cavity at the scale of the micron and uniquely based on the absence of thermal equilibrium.

\begin{acknowledgments}
The authors thank D. Felbacq and B. Guizal for fruitful and stimulating discussions. The authors acknowledge financial support from the Julian Schwinger Foundation and from the LabEx NUMEV.
\end{acknowledgments}

\end{document}